# COVID-Town: An Integrated Economic-Epidemiological Agent-Based Model

Patrick Mellacher[1]

**This version: 10th of November**

**Abstract**: I develop a novel macroeconomic epidemiological agent-based model to study the impact of the COVID-19 pandemic under varying policy scenarios. Agents differ with regard to their profession, family status and age and interact with other agents at home, work or during leisure activities. The model allows to implement and test actually used or counterfactual policies such as closing schools or the leisure industry explicitly in the model in order to explore their impact on the spread of the virus, and their economic consequences. The model is calibrated with German statistical data on time use, demography, households, firm demography, employment, company profits and wages. I set up a baseline scenario based on the German containment policies and fit the epidemiological parameters of the simulation to the observed German death curve and an estimated infection curve of the first COVID-19 wave. My model suggests that by acting one week later, the death toll of the first wave in Germany would have been 180% higher, whereas it would have been 60% lower, if the policies had been enacted a week earlier. I finally discuss two stylized fiscal policy scenarios: procyclical (zero-deficit) and anticyclical fiscal policy. In the zero-deficit scenario a vicious circle emerges, in which the economic recession spreads from the high-interaction leisure industry to the rest of the economy. Even after eliminating the virus and lifting the restrictions, the economic recovery is incomplete. Anticyclical fiscal policy on the other hand limits the economic losses and allows for a V-shaped recovery, but does not increase the number of deaths. These results suggest that an optimal response to the pandemic aiming at containment or "holding out for a vaccine" combines early introduction of containment measures to keep the number of infected low with expansionary fiscal policy to keep output in lower risk sectors high.

**Funding:** This research did not receive any specific grant from funding agencies in the public, commercial, or not-for-profit sectors.

## 1 Introduction

Agent-based modeling was used to explain (features of) the spread of the SARS-COV2 virus among others by the Washington Post (2020) and Paul Romer (Romer 2020a, Romer 2020b). It rose so much to prominence that one may nearly get the impression that "We are all agent-based modelers now", although in reality S(E)IR compartment models are still widely used by epidemiologists, who want to predict how the virus spreads and how the spread is affected by various policies (see e.g. Walker et al. 2020), and rule the roost among economists who develop coupled epidemiological-economic models (e.g. Acemoglu et al. 2020, Eichenbaum et al. 2020).

[1]University of Graz, Graz Schumpeter Centre, patrick.mellacher@uni-graz.at
Universitätsstraße 15/F, 8010 Graz, Austria

Both simple ABMs, as well as S(E)IR compartment models treat an epidemic as a process, in which agents differ with respect to their epidemiological state (i.e. susceptible, (exposed), infected and recovered/immune) and randomly meet each other to eventually spread the disease. In simple ABMs, this is modeled with agents who roam freely in an undefined space. SIR compartment models do not model each agent explicitly. Instead, they assume homogenous mixing, which means that each agent (of one epidemiological state) has the same probability to meet another agent. Due to the homogenous mixing assumption, SIR models can be run just as efficiently for a large simulated population as for a small one. This is a great asset during a pandemic, since forecasts need to be provided quickly in order to advise policies. Simple ABMs on the other hand are powerful in graphically explaining key features of an epidemic like exponential growth to a wider audience, and at the same time incorporate an important stylized fact, namely the need to physically meet each other to spread the disease. These models thus help to raise awareness and the level of understanding.

While both types of well-known models clearly have their merits, they are limited in explaining and incorporating other important stylized facts of the COVID-19 pandemic. Random encounters, such as by two people who go for a random walk and meet each other at the sidewalk, seem to play a negligible role in the spread of the virus.[2] Furthermore, people typically do not move randomly, but consciously and tend to stay at some places for a longer time (e.g. at home, at work, at school, at a club or hospital etc.). The risk of meeting an infected person is thus much greater for a barkeeper in a big club and even more for a doctor in a hospital, than it is for an old lady who largely stays home alone. This fact likely has a nontrivial impact on the development of a pandemic. Think of a leisure contact that happens on a monthly basis and may – or may not – cause an infection, which then spreads fast among household or work contacts that happen on a daily basis.

A more recent generation of agent-based models distinguishes between different settings of social contacts, most commonly work, leisure and household contacts. The contacts in these settings differ with regard to regularity and/or transmissibility. Vermeulen et al. (2020) present an "An agent-based policy laboratory for COVID-19 containment strategies", in which leisure, work and household locations and shops are modeled as boxes. Agents move randomly within these boxes, as in the simple ABMs by e.g. Romer, but also consciously between boxes at specific times of the simulated day. Since their model is interactive, it allows users to explore complexities involved in decisions such as putting a particular location under quarantine. Bicher et al. (2020) and Kerr et al. (2020) develop ABMs, in which agents have regular social contacts at their households, workplaces and schools, but random contacts during leisure.

Most epidemiological ABMs focus on the epidemiological aspects of the pandemic and thus do not integrate the socioeconomic effects of the various lockdown measures, although there are some notable exceptions that I will discuss below. Some economic ABMs, on the other hand, are employed to study the economic impact of containment measures, but do not integrate epidemiological mechanics into their model: Sharma et al. (2020) discuss various recovery scenarios for the Post-COVID recovery. Inoue and Todo (2020) analyze how a lockdown in Tokyo affects output in Tokyo directly and other Japanese regions indirectly. Poledna et al. (2020) use a macroeconomic ABM of the Austrian economy to study the medium-term economic impact of the lockdown measures.

---

[2] See e.g. Böhmer et al. 2020, who show in their case study of an outbreak in Bavaria involving 17 cases that "low-risk" contacts, i.e. face-to-face contacts of less than 15 minutes did not cause any infection.

Studying the interplay between the economy and the pandemic is thus largely left to SIR models so far, which have been extensively used since a highly influential article by Atkeson (2020). These models focus mostly on abstract "optimal lockdown" strategies (e.g. Acemoglu et al. 2020, Eichenbaum et al. 2020) and allow to explore the trade-off between economic performance and the spread of the disease. However, due to the limited amount of heterogeneity that can be included into an SIR model while keeping its tractability this is only possible in a highly stylized way, which potentially results in biased policy advice.

SIR compartment models are governed by the effective reproduction number $R_t$, which denotes the number of people that each infected person infects. They suggest two viable strategies to deal with the COVID-19 pandemic in the long term: One is to achieve "herd immunity" or "partial herd immunity" (e.g. Acemoglu et al. 2020, Alvarez et al. 2020). A version of this strategy, where the total number of infections is unchanged, but the epidemic is mitigated, such that the peak of infections is reduced and the capacities of the health care system are not exhausted (or at least not too much) was popularized as "flattening the curve". The other alternative, which became famous as "The Hammer and the Dance" (Pueyo 2020), but is called suppression in the scientific discourse (Ferguson et al. 2020) aims to quickly reduce the number of infections to a manageable size (the "hammer") and then to keep the reproduction number at around or below 1 (the "dance") and holdout until a vaccine or a cure is developed. The analysis by Atkeson (2020) suggests that such an approach is economically extremely expensive.

In classical SIR models, it is not possible to eliminate the virus via containment measures before reaching herd immunity, except if the number of contacts is brought down to zero, a special case that would typically be excluded by correctly claiming that e.g. some basic production is necessary to allow society to survive (e.g. Acemoglu et al. 2020). Piguillem and Shi (2020) allow for an elimination of the virus in their model, by assuming that once less than one person is infected, the virus is eliminated. In a standard SIR model, fractions of an individual would still continue to spread the virus, allowing for a resurgence once the restrictions are lifted. In an agent-based model, the number of agents is discrete and such a scenario is thus by design impossible.

By emphasizing the existence and heterogeneity of various sources of infection, agent-based models may be able to identify not only those policies, which seem to be most promising from an epidemiological point of view, but also from a socioeconomic one. Closing schools and childcare facilities, for instance, significantly reduces the number of social contacts, but places a high burden on parents, especially with younger kids, who then have to take care of the children and either cannot go to work or have to juggle with childcare, distance learning and teleworking. Closing down those industries, which seem to be most problematic in the spread of the virus (e.g. the leisure industry) will place a disproportionally high burden on some groups of society. But if it is enough to save economic activity in other industries, it may be possible to compensate the losses.

The model presented in this paper aims to be a first step in building an epidemiological agent-based model along the lines of state-of-the-art models of the field by e.g. Aleta et al. (2020), Bicher et al. (2020) and Kerr et al. (2020) which is compatible with the macroeconomic agent-based literature.[3] These epidemiological models focus on the spread of the disease via various social networks in contrast to homogenous mixing between as seen in standard SIR compartment models. These models distinguish between social contacts in schools, workplaces, households and during leisure time (sometimes called "community" contacts).

This model is an integrated model, since it allows to study the effects of policies on economic and epidemiological outcomes at the same time. In contrast to SIR compartment models, this model also allows to evaluate the impact of explicitly modeled social network interventions (e.g. school closures) on public health and the economy.

Macroeconomic integrated assessment models have already been built and fruitfully employed to analyze the interplay between the long-term evolution of the economy and climate change (Lamperti et al. 2018a, 2018b; Hötte 2019; Rengs et al. 2020). In contrast to climate change, however, outcomes of the COVID crisis are also highly relevant for the ultra-short run, i.e. days or fractions of days. Macroeconomic ABMs are, however, typically built to capture medium- to long-run trends. For instance, in the "benchmark" model by Caiani et al. (2016), each period is calibrated to match one quarter. In the Keynes + Schumpeter model by Dosi et al. (2010), the average GDP growth rate per period is 2.5%, so each period matches approximately one year. The most detailed picture is given by the EURACE@Unibi model (see Dawid et al. 2019 for a full documentation of the model), where each time step represents one business day. Basurto et al. (2020) develop an economic epidemiological model building on this model and an enhanced SIR framework.[4]

Because of the perceived difficulties of immediately combining a full-fledged economic model with a full-fledged epidemiological model, I chose the opposite approach and built a new model. The economic part of this model is highly simplified in many aspects. This approach was chosen in order to reduce the time necessary to build, verify, calibrate and validate the model. However, a) the present model can be extended in numerous ways in order to highlight other aspects and explore the interrelation of the economy and the pandemic with increased complexity, and b) this model may also serve as a starting point to carve out an epidemiological module that can be integrated into other macroeconomic agent-based models.

I calibrate this model to German data, since high-quality data on the German epidemiological curve is available due to a) the relatively high number of tests and b) the nowcasting table published by the Robert Koch Institut (2020b), which estimates the true date of infection based on multiple imputation techniques. After successfully calibrating the epidemiological parameters, however, they can then be used to investigate other countries in the future.

---

[3] See Dawid and Delli Gatti (2018) for a recent overview of the macroeconomic ABM literature.
[4] A more detailed account of the model and its differences to COVID-Town is given below.

Four other epidemiological agent-based models which include economic models or variables were developed so far: Vermeulen et al. (2020) integrate labor supply as an outcome variable of their "policy laboratory". My model differs from it by providing a much more detailed account of the economy. Also, labor supply in their model does not seem to depend on whether schools are open or not, whereas in my model, a caregiver has to take care of young children and cannot leave their home for work. On the other hand, the model by Vermeulen et al. (2020) is much more detailed than mine when it comes to the representation of time, since they simulate minutes, whereas I only split each day in three phases. Silva et al. (2020) present a model to study the effects of various counterfactual containment scenarios on a virtual economy representing Brazil. Among others, my model features schools and different types of work places to account for the policy responses observed in Europe. Their model, on the other hand, features stylized facts which are particularly important for the case of Brazil and similar countries like homelessness and different types of social stratums. The ASSOCC model introduced by Dignum et al. (2020) includes an epidemiological, a cultural, a needs model, an epistemic model, an economic model and a transport model. A full documentation of the ASSOCC model is unavailable, most probably because of its massive scale indicated by the source code and various shorter descriptions. Their model seems very promising in studying the results which emerge from the interplay of the various models. COVID-Town differs from the ASSOCC model and the other two models discussed especially with regard to a) **the economy**: COVID-Town emphasizes the heterogeneity of industries and of different types of workers as explicated in more detail below, b) **calibration**: While most models use statistical data to calibrate age and household structures, I additionally employ data on time use, employment, wages, firm demography and company profits to set up the artificial economy, and c) **validation:** I set up a baseline scenario using statistical data that is validated quantitatively. This would be almost impossible for the other three models discussed, at least for a country like Germany, because they operate with a low number of agents (a few hundred each) that cannot account for the small fraction of people infected during the first wave in Germany and similar countries.

Finally, Basurto et al. (2020) develop a model inspired by the EURACE@Unibi macroeconomic agent-based model by Dawid et al. (2019). They base the spread of the virus on a "standard SIRD model" (Basurto et al. 2020, p. 3) with three different types of contacts: workplace, shopping and private contacts. While workplace contacts follow a social network pattern in which workers meet their co-workers, the other two channels are random (between groups).[5] This model is closest to mine, since it also does not only feature different jobs, but also different professions, differentiates between various branches of the economy and is calibrated to and validated with empirical data on infections and deaths (in contrast to the other three models discussed). Building on a well-tested rich macroeconomic framework, they model e.g. the firm's pricing decision and evolution of market share in a more detailed way (compared to the consumption goods industry in COVID-Town) and emphasize two explicitly modeled economic policies: short-term work and firm bailout policies. Relative to their model, my model differs and contributes to the literature model especially in the following four aspects: a) **virus transmission:** In COVID-Town, social networks do not only govern work place contacts, but also leisure and household contacts.[6] In addition to that, I also model contacts between patients and health care workers as well as contacts between service workers and customers, both of which seemed to have played a crucial role in the pandemic e.g. in Ischgl (Politico 2020) and Bergamo. This enables me to implement containment policies such as the isolation of family members and colleagues and study their combined effects on public health and the economy. b) **agent heterogeneity**: My model features eight types of archetypal human agents who differ distinctly with regard to their economic and epidemiological characteristics (e.g. Do they have contact with customers? Can they work from home?) described in table 1. This setup can help to identify those who are pivotal to the spread of the virus, but also those who suffer the most from the containment measures. c) **leisure (industry)**: In COVID-Town, agents have age-specific heterogeneous preferences for spending their leisure (staying at home, meeting friends, visiting commercial or non-commercial leisure facilities). Leisure time is also explicitly modeled, which means e.g. that most agents engage in more leisure activities over the weekend and unemployed workers have more leisure time than employed ones. This allows me to e.g. study the epidemiological effects of an increase in unemployment (which translate to a decrease in work contacts, but an increase in leisure contacts), but also e.g. the combined effects of a Swedish-style encouragement of self-restraint in which agents choose to stay at home voluntarily more often vs. a strict lockdown-policy in which the leisure industry is closed or a curfew is announced, and d) **care work**: in my model, school closures force parents of young children to work from home (or abstain from working, if they are unable to telework). My model thus allows me to shed light on the trade-off between the economic and the epidemiological impact of school closures, which did not receive attention so far in the model-based literature, but plays an important role in the public discourse.

---

[5] Households in their model meet other households on shopping days, which are randomly chosen in every week and thus akin to homogenous mixing, whereas private meetings follow an age-specific homogenous mixing approach, similar to the multi-risk SIR model by Acemoglu et al. (2020).

This paper offers two main contributions. The first one is the development of a novel heterogeneous agent-heterogeneous mixing integrated assessment model of the Covid-19 pandemic that can be calibrated with German statistical data on time use, demography, business demography, household composition, employment and other economic statistics to fit both observed German deaths and the estimated infection curve and account for a large number of stylized facts on the Covid-19 crisis. This model allows to shed light on the complex interplay triggered by explicit containment policies such as closing schools or the leisure industry in the following interconnected dimensions:

1. **The economic dimension**: How are output and employment affected?
2. **The medical dimension**: How many people become infected and die? Are the containment policies sufficient to avoid an overloaded health care system?
3. **Care work**: Closing schools may help to reduce the spread of the infection, but these policies drastically increase the care work that must be done by the families. How does this feed back to the economic dimension in terms of a reduction in labor supply and less efficient teleworking hours?
4. **Leisure**: How do the containment policies influence the ability of the agents to spend their leisure as planned?

---

[6] In the model by Basurto et al. (2020) all households consist of a single person. Household transmission (identified by as the most common source of infection by e.g. Lee et al. 2020) is thus not included in their model. Modeling (different types of) households explicitly enables me to study e.g. to study how pensioners living in intergenerational households are affected by the crisis relative to pensioners living in retirement homes or pensioners living alone. On the other hand, my model does not feature shopping contacts.

The second main contribution concerns the economy in this model: It features eight types of human agents, three branches of the private industry (offices, factories and commercial leisure facilities) and three branches of government activity (schools, hospitals and other government purchases). Since economic activity in this model is heterogeneous with regard to its incorporated social contacts, containment policies can be targeted at those types of economic activity that involve the most interactions (i.e. leisure facilities) or where it is easily possible to reduce the number of social contacts (i.e. white-collar workers working from home instead of the office). While epidemiologically extremely useful, closing down commercial leisure facilities like clubs and restaurants can have a devastating economic impact through ripple effects that drag down private and public consumption, especially if fiscal policy is constrained by a zero-deficit clause. This effect is larger than the "Keynesian supply shock" suggested by Guerrieri et al. (2020), since it also involves shrinking public expenditure. My model suggests that anticyclical fiscal policy is crucial in combatting the economic recession and that it is thus highly important to enable countries affected by high debts to ramp up their public expenditure, too.

In addition to that the paper offers two smaller contributions:

1.) It contributes to the epidemiological ABM literature by introducing stylized facts especially important in this pandemic, but not yet included by the main ABMs in the field: retirement homes (included in the most recent model by Bicher et al. 2020), hospitals and commercial vs. non-commercial leisure facilities as sources of transmission.
2.) This paper furthermore contributes to the economic literature by introducing an agent-based analysis of leisure and the leisure industry. This is particularly useful in the COVID-pandemic, since a) leisure is an important source of transmission, which cannot fully be captured by SIR compartment models due to the heterogeneous mixing observed in this sphere of the real world and b) conflicting interests revolve around the epidemiological and economic role of the leisure industry.

The rest of the paper is organized in the following way: Section 2 gives a detailed account of the model. Section 3 describes its calibration and section 4 reports the policies which are implemented and tested for this paper. Section 5 and 6 discuss model verification and validation respectively. Section 7 describes the results of counterfactual policy scenarios. Section 8 discusses some of the limitations of the model and how it can be improved in the future. Finally, section 9 concludes.

**2 The Model**

The basic idea of this paper is to create an artificial town – COVID-Town – which is exposed to a COVID-19 outbreak and various policies to combat the spread of the virus. This town is populated by agents (described in 2.1), who behave according to boundedly rational rules and are connected in a social network (as shown in 2.2). It is calibrated to a representative empirical German town using various statistical sources (see section 3). In this model, infections take place in explicitly modeled places. Such a place belongs to at least one of three spheres: home, work (including school) and leisure. A place may be a leisure location for some and a work location for others. Agents are connected to these places via social networks (as described in 2.3). COVID-Town aims to represent the empirics of these spheres, albeit in a more or less stylized way in order to a) calibrate the model by setting up a validated baseline scenario capturing the empirical dynamics (see section 3 and 6) and b) deviate from this baseline using what-if policy scenarios (section 7).

The model is discrete-time, where each simulated day is split into three phases that – for a typical employee on a weekday – is devoted to each sphere. During each phase, a sequence of events shown in figure 2 is computed.

The model is implemented using Netlogo (Wilensky 1999) and analyzed using the ggplot2 package (Wickham 2016) for the programming language R (R Core Team 2018). Since a large number of stochastic processes are involved in setting up and running the model, I rely on Monte Carlo simulations to analyze the results, i.e. I run a large number of experiments on the same parameters in order to quantitatively estimate the overall dynamics of the system.

*2.1 Agents and their characteristics*

This model follows a tradition of macroeconomic agent-based models, in which the population is divided into various types of workers and/or social classes (Caiani et al. 2019, Rengs and Scholz-Wäckerle 2019, Mellacher and Scheuer 2020). This model features eight types of agents, who differ with regard to their role in the economy, their interactions at work, their teleworking ability, as well as their age span (see table 1). Their characteristics are stylized representations of the real world. For instance, not everybody from 0 to 19 actually goes to school or a child care facility. These stylizations are necessary, however, in order to see not only the trees, but also the forest.

**Table 1: Agent types and their characteristics**

| Agent type | Age | Economic Role | Interactions at work |
|---|---|---|---|
| Child/Young Person | 0-19 | Goes to school. | Regular interactions with other children and teachers |
| White Collar Worker | 20-64 | Works in an office producing white collar services. May work from home. | Regular interactions with co-workers |
| Blue Collar Worker | 20-64 | Works in a factory producing consumption goods. | Regular interactions with co-workers |
| Service Worker | 20-64 | Works in a commercial leisure facility. | Regular interactions with co-workers + random interactions with guests |
| Health Care Worker | 20-64 | Works in a hospital. | Regular interactions with co-workers + random interactions with patients |

| Teacher | 20-64 | Teaches children. | Regular interactions with children and other teachers |
|---|---|---|---|
| Pensioner | 65+ | - | - |
| Firm owner | 20+ | Receives rents from his/her offices, factories and commercial leisure facilities. | - |

*2.2 Social interactions*

Social contacts are at the center of any epidemiological investigation. As sketched in figure 1, interactions in this model belong to one of three spheres – home, work or leisure – and differ with regard to their frequency.

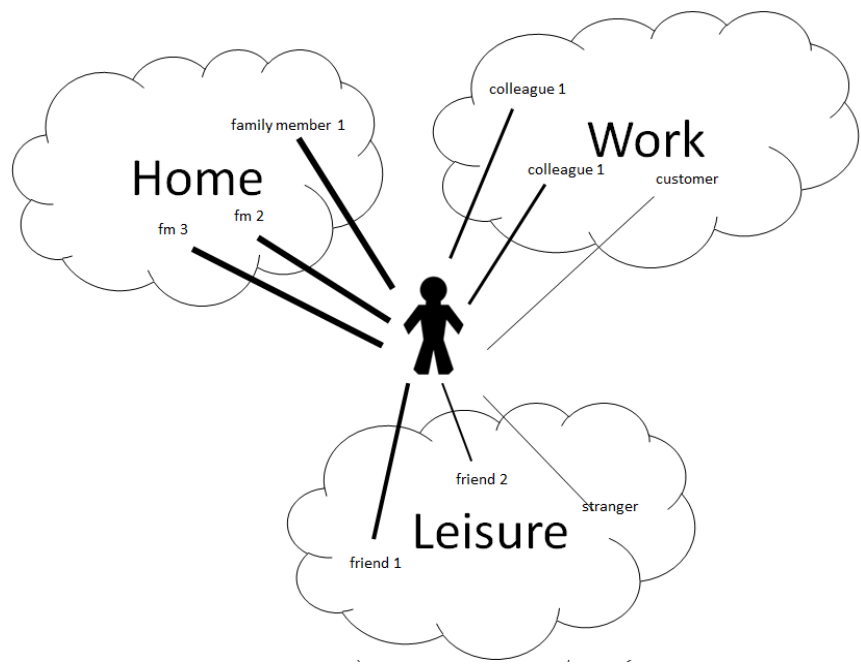

**Figure 1: Social contacts of a person interacting in the different spheres. Thicker lines denote higher frequencies of contact, i.e. agents meet relatively often.**

*2.2.1 Home*

Every human agent in the model is assigned to a household (or retirement home). The household composition is taken from Statistisches Bundesamt (2019b). Household members meet at least during night on every day (except if one of them has to do a night shift). Interactions at home are thus very frequent, which makes it likely that an infected family member infects the others. Agents may also choose to stay at home during their leisure phase and have to stay at home during their work phase, if they are isolated or their workplace is closed. Agents who had to do a night shift will stay at home during the first phase of the next day.

*2.2.2 Work*

Every employed worker (based on labor market statistics from Bundesagentur für Arbeit 2020a) and child (all children and young people are assumed go to school) is assigned to a work place / school. Interactions at work happen less frequently than at home, since workers and children are assumed to go to work five times a week. Since I assume that each individual only has at maximum $\gamma$ social contacts per period, the agents do not necessarily meet all their co-workers/customers.

In a small company without customer contact, these interactions are thus still predictable, since they involve a relatively constant number of people. A random element is introduced for workers who face customers, patients or students. In this model, this is true for teachers, health care and service workers. Interactions with customers are more sporadic and less predictable – they are thus much more akin to the "homogenous mixing" assumption of standard SIR models than interactions with coworkers and at home.

Children, teachers, white-collar workers and blue-collar workers are assumed to have one work phase on each day from Monday until Friday. Health care workers are on the other hand assigned to shifts to make sure that hospitals are staffed 24/7. Service workers are also assigned to shifts at commercial leisure facilities, but the latter are only open for a total of 14 shifts (i.e. the possible two leisure phases per day multiplied by the number of days per week). Since the shifts on the weekend and in phase 2 of each day (i.e. afternoon/evening) are attracting more customers, leisure facilities aim to have twice as many workers during these shifts as during off-peak shifts.

*2.2.3 Leisure*

Human agents, who are at least 10 years old have up to two leisure phases per day: one on a work day, two on a day off (i.e. on the weekend for children and all workers except service and health care workers). For each leisure phase, agents make plans on how to spend their leisure. The following activities are available: meet a friend, visit a non-commercial leisure facility (representing e.g. parks), visit a commercial leisure facility (e.g. bars) or stay at home.

Agents are connected to their friends and their preferred leisure facilities in a social network via edges. These edges are weighted randomly to account for the fact that some friends and places are more important to us than others. The weights, as well as their preference for staying at home, are drawn from a normal distribution with the age-specific mean $\mu_{a,b}$, where $a$ denotes the age and $b$ the activity, as well as the standard deviation $\sigma$ for each agent-activity pair. The preferences are calibrated to match on average the observed preferences from the German time use study (Statistisches Bundesamt 2015).

These preferences are then used to set up a leisure plan, i.e. to decide, which activity is most-preferred, second-most preferred and so on. The preferences are described by utility gained from various activities, but the agents do not maximize their utility. Instead, they decide on their leisure plan using a stochastic boundedly rational process.[7]

More specifically, agents may gain utility u from

1.) Meeting a friend, which is described by the weight of the edge between the two agents $u_{j,j^*}^{friend}$. This weight is drawn in the beginning of the simulation and then remains constant.
2.) Going to a non-commercial leisure facility, which is equal to the weight of the edge $u_{j,z^*}^{nc}$ multiplied with the facility-specific attractiveness $a_{z^*}$.
3.) Going to a commercial leisure facility, which is equal to the weight of the edge $u_{j,z^{**}}^{c}$ multiplied with the facility-specific attractiveness $a_{z^{**}}$, multiplied with the parameter $\kappa$ that accounts for the fact that the observed time spent at commercial leisure facilities is likely lower than the preferred one if there were no monetary constraints.
4.) Staying at home $u_{j,t}^{home}$.

In the first step, agents sum up the utility they would gain from doing all activities simultaneously to $f_{j,t}$, where j denotes the individual and t the time step.[8]

$$f_{j,t} = \sum_{j^*} u_{j,j^*}^{friend} + \sum_{z^*} u_{j,z^*}^{nc} a_{z^*} + \sum_{z^{**}} u_{j,z^{**}}^{c} a_{z^{**}} + u_{j,t}^{home}$$

A draw from a uniform distribution from 0 to (excluding) $f_{j,t}$ then decides which activity is most-preferred. A second draw determines the second-most preferred activity etc. using the following rule: if the drawn number is lower than $u_{j,j'}^{friend}$, where $j'$ denotes the first friend, the agent wants to meet $j'$. If the number is higher than $u_{j,j'}^{friend}$, but lower than $u_{j,j''}^{friend}$, where $j''$ denotes the second friend, the agent wants to meet the second friend and so on. A maximum leisure plan length is set for all individuals.

I then employ a matching procedure to carry out the plans: First, all agents who want to go to leisure facilities (or stay at home) try to execute their plans. After this step, only agents are left who plan to meet their friends as their most-preferred activity and they try to do so. This is only possible if their friends are not isolated, in the hospital, or meeting somebody else. If all agents, who wanted to do so, tried to meet their friends, again every agent who wants to visit a leisure facility or stay at home tries to carry out its plans and so on.

---

[7] Since each agent can only engage in a single activity per leisure period, utility maximization with constant utilities would not make much sense as agents would always choose the same activity. In order to incorporate a utility maximization framework into this discrete-time, explicit place model, one would have to devise a much more sophisticated utility function, in which the utility from each activity depends on the past activities. While such an approach might be intriguing theoretically, its practical added value for this study would be little, since the goal of the leisure plan mechanism is to mimic the empirical leisure behavior of Germans. The mechanism described here is both simple and able to fit the empirics, but can also account for the stylized fact that observed choices do not necessarily equal actual preferences.

[8] The time step index is necessary, because the home preference is redrawn from the normal distribution at the beginning of each leisure phase, to account for the fact that we simply want to stay at home (or go out) on some days. The home preference is also multiplied with a home preference multiplier. The default value for this multiplier is 1, but it may be altered as part of a scenario in order to account for voluntary/encouraged social distancing.

If all plans fail, e.g. because the preferred friends do not want to meet, or are unable to do so because the preferred location is under quarantine, their friends are isolated etc. the agent stays at home. Again, agents meet $\gamma$ (or the total number of other agents at the current location, if it is smaller than $\gamma$) other agents.

## 2.3 Sequence of Events

In each time step of this model, the sequence of events shown in figure 2 is computed. Some activities only take place in specific phases and others only on specific days. Agents move between locations, work and interact according to their shift schedule. The infection module is described in 2.4 and a detailed account of the economy is given in 2.5.

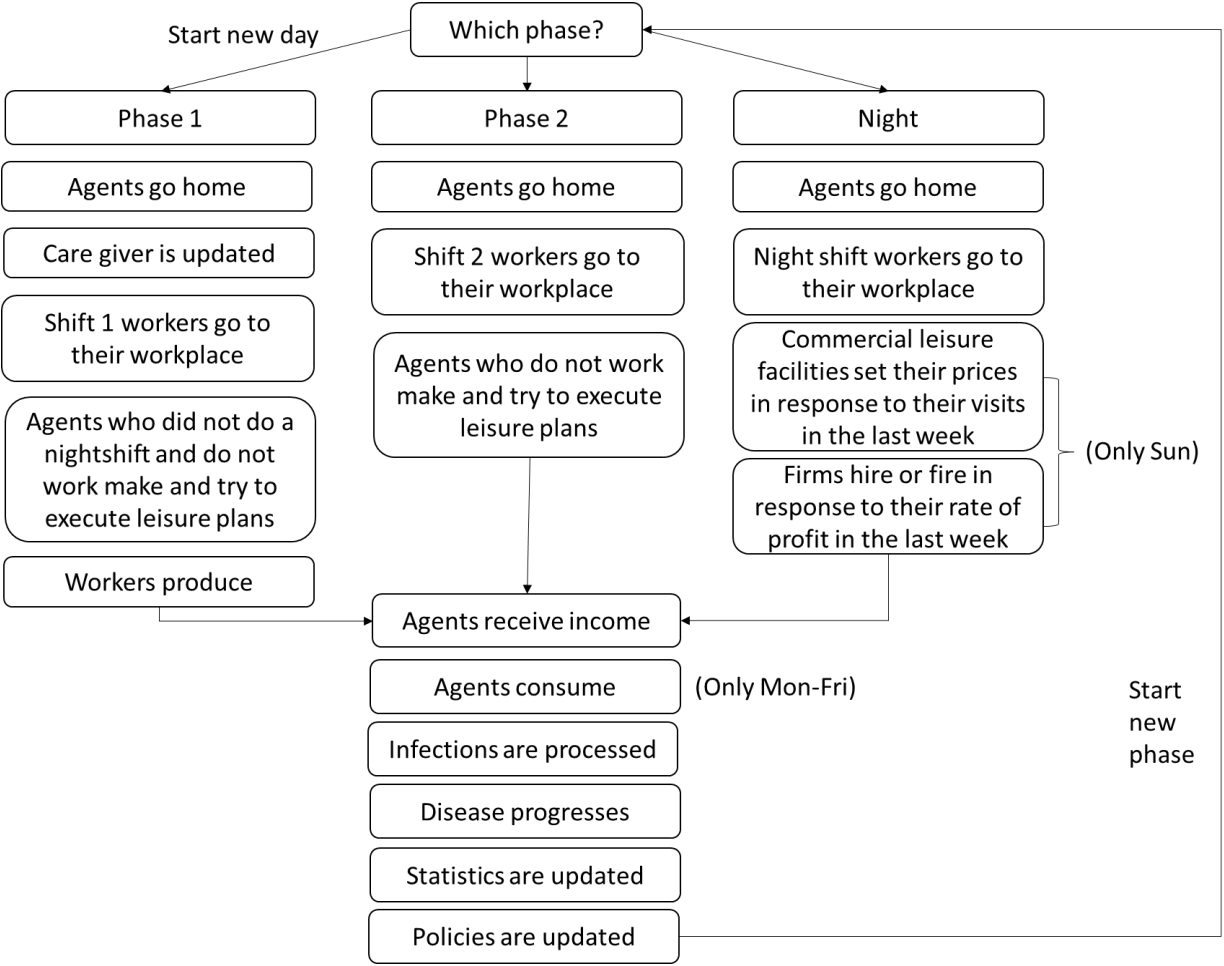

**Figure 2: Sequence of events**

## 2.4 Modelling the infection

### 2.4.1 Infections

In each phase, each infected agent is assumed to come into contact with (up to) $\gamma$ other agents at the same location. Each contact with a susceptible agent infects the latter with the probability $\beta x_{i,t} \pi_p$, where $\beta$ denotes the baseline infection probability, which is altered by an overcrowding variable $x_{i,t}$ and a hygiene parameter $\pi_p$. $x_{i,t}$ is only calculated for leisure facilities by dividing the number of agents at a certain location $n_{i,t}$ by its capacity ($\vartheta_1^{II}$ for commercial and $\vartheta_2^{II}$ for non-commercial leisure facilities). The overcrowding variable is thus determined endogenously for leisure facilities. For all other locations it is assumed to be 1. $\pi_p$ on the other hand, denotes a hygienic parameter that represents increased hygiene and use of personal protective equipment at locations of type p (e.g. hospitals). This parameter is exogenous and can be altered as part of a strategy (e.g. increased hygienic standards at hospitals in order to avoid infections of hospital staff).

A special rule is in place for children who are at school: they are assumed to come into contact with $\gamma - 1$ other children of their class and one additional random person that is present at their school.

*2.4.2 Progression of the disease*

The progression of the disease is based on a state-of-the-art SEIR model proposed by the Imperial College (2020) as shown in figure 2. Whenever an agent becomes infected, a severity level $\sigma_j$ from 0 to 1 is drawn from a uniform distribution. A severity level of 0.5 is the median and means that 50% of the infections are more severe and 50% less severe. As it can be seen from table 2, only 0.1% of the infected belonging to the group of 0 to 4-year-old must be hospitalized at some point compared to 18% of the infected of 80 years and above. The effects of this severity level are thus age-specific.

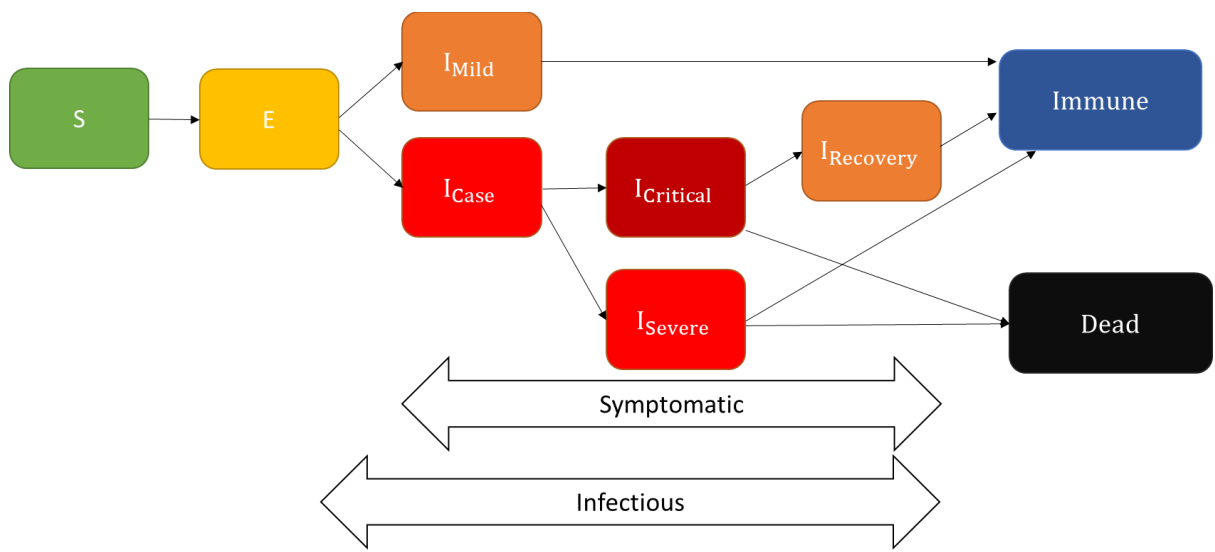

**Figure 3: Progession of the disease, based on Imperial College (2020).**

Each agent gets randomly assigned to one age group that corresponds to the respective age span shown in table 1. The probability distribution of the age groups is given by German demographic data (Statistisches Bundesamt 2019). The age group then governs the medical characteristics of a person (see table 2).

**Table 2: Age groups and their medical characteristics**

| Age group | Proportion of infected hospitalized $\rho_a^I$ | Proportion of hospitalized | Proportion of non-critical care cases dying $\rho_a^{III}$ | Initial population |
|---|---|---|---|---|

|  |  | cases requiring critical care $\rho_a^{II}$ |  |  |
| --- | --- | --- | --- | --- |
| 0 to 4 (0) | 0.001 | 0.050 | 0.013 | 0.0473 |
| 5 to 9 (1) | 0.001 | 0.050 | 0.013 | 0.04411 |
| 10 to 14 (2) | 0.001 | 0.050 | 0.013 | 0.04459 |
| 15 to 19 (3) | 0.002 | 0.050 | 0.013 | 0.04822 |
| 20 to 24 (4) | 0.005 | 0.050 | 0.013 | 0.0555 |
| 25 to 29 (5) | 0.010 | 0.050 | 0.013 | 0.06256 |
| 30 to 34 (6) | 0.016 | 0.050 | 0.013 | 0.06515 |
| 35 to 39 (7) | 0.023 | 0.053 | 0.013 | 0.06309 |
| 40 to 44 (8) | 0.029 | 0.060 | 0.015 | 0.05832 |
| 45 to 49 (9) | 0.039 | 0.075 | 0.019 | 0.06727 |
| 50 to 54 (10) | 0.058 | 0.104 | 0.027 | 0.08282 |
| 55 to 59 (11) | 0.072 | 0.149 | 0.042 | 0.07948 |
| 60 to 64 (12) | 0.102 | 0.224 | 0.069 | 0.06618 |
| 65 to 69 (13) | 0.117 | 0.307 | 0.105 | 0.05792 |
| 70 to 74 (14) | 0.146 | 0.386 | 0.149 | 0.04332 |
| 75 to 79 (15) | 0.177 | 0.461 | 0.203 | 0.04925 |
| 80+ (16) | 0.180 | 0.709 | 0.580 | 0.05574 |

**Source**: Medical characteristics from Imperial College (2020), based on Verity et al. (2020). Initial population from Statistisches Bundesamt (2019a). Please note that these characteristics may be subject to mutations and/or better treatments.

If $\sigma_j$ is higher than or equal to the proportion of infected hospitalized, this person must be hospitalized after a certain number of periods (otherwise, the infection will take a mild course) described in section 3. If $\sigma_j$ is furthermore higher than or equal to (1 - proportion of infected hospitalized $\rho_a^I$ * proportion of hospitalized cases requiring critical care $\rho_a^{II}$), this person will need an intensive care unit to have a chance to survive in the future (otherwise it will be a severe case in need of a hospital bed).

Infections which take a severe or a critical course may cause a death in the end. More precisely, following the Imperial College (2020), it is assumed that 50% of all people who are in critical care die. Thus, if $\sigma_j \geq 1 - 0.5\rho_a^I\rho_a^{II}$, an ICU patient is set to die, other ICU patients will survive. If an agent who is a critical case does not receive intensive care it is assumed to instantly die. Some severely infected who get a hospital bed may also end up dead, although their proportion $\rho_a^{III}$ varies greatly between the age groups. Technically, this applies to those infected, who have a $\sigma_j \geq 1 - \rho_a^I\rho_a^{II} - \rho_a^I(1-\rho_a^{II})\rho_a^{III}$, i.e. those severe cases with the highest severity level. If an agent who is a severe case does not get a hospital bed it dies with a flat probability of 60% (again, following Imperial College 2020). These agents are to be found *mutatis mutandis*, i.e. those 60% of the severe cases with the highest $\sigma_j$.

It is assumed that the severity level is unknown to the agents and hospitals can thus not choose e.g. to provide intensive care only to those who would survive it. Hospitals also do not use triage to identify, who they admit, but allow admissions on a first come, first serve basis (as long as capacity is available).

*2.4.3 Deaths*

If an agent dies, an heir is randomly determined from the population of 20+ year old people. This heir receives all remaining funds (and firms, if the dead agent was a firm owner). If the dead agent was the last possible care giver to a child, this child is randomly assigned to a new family.

*2.5 The Economy*

Since each time step of the simulation represents one third of a day, the virtual economy in this model does not only have to represent the real German one – albeit in a stylized way – and but should in absence of the virus also not produce drastic changes endogenously that would a) be unexpected in the real economy in such short periods of time and b) make it harder to dissect the changes resulting from the virus and the containment measures from the changes that arise endogenously.

In order to solve this challenge, the economy devised in this paper is rather simple. Nevertheless, this simple model can be extended in numerous ways to incorporate important stylized facts and processes studied in detail by macroeconomic ABMs like capital goods, innovation, explicit market interactions at the consumption goods market (e.g. Dosi et al. 2010, Caiani et al. 2016, Dawid et al. 2019) in the future and it would be highly interesting to study how the dynamics described by my model are affected by these stylized facts.

Figure 4 gives an overview of the agents and markets in the economy as well as their economic relations. The exact mechanisms implemented are described in detail in the following subsections.

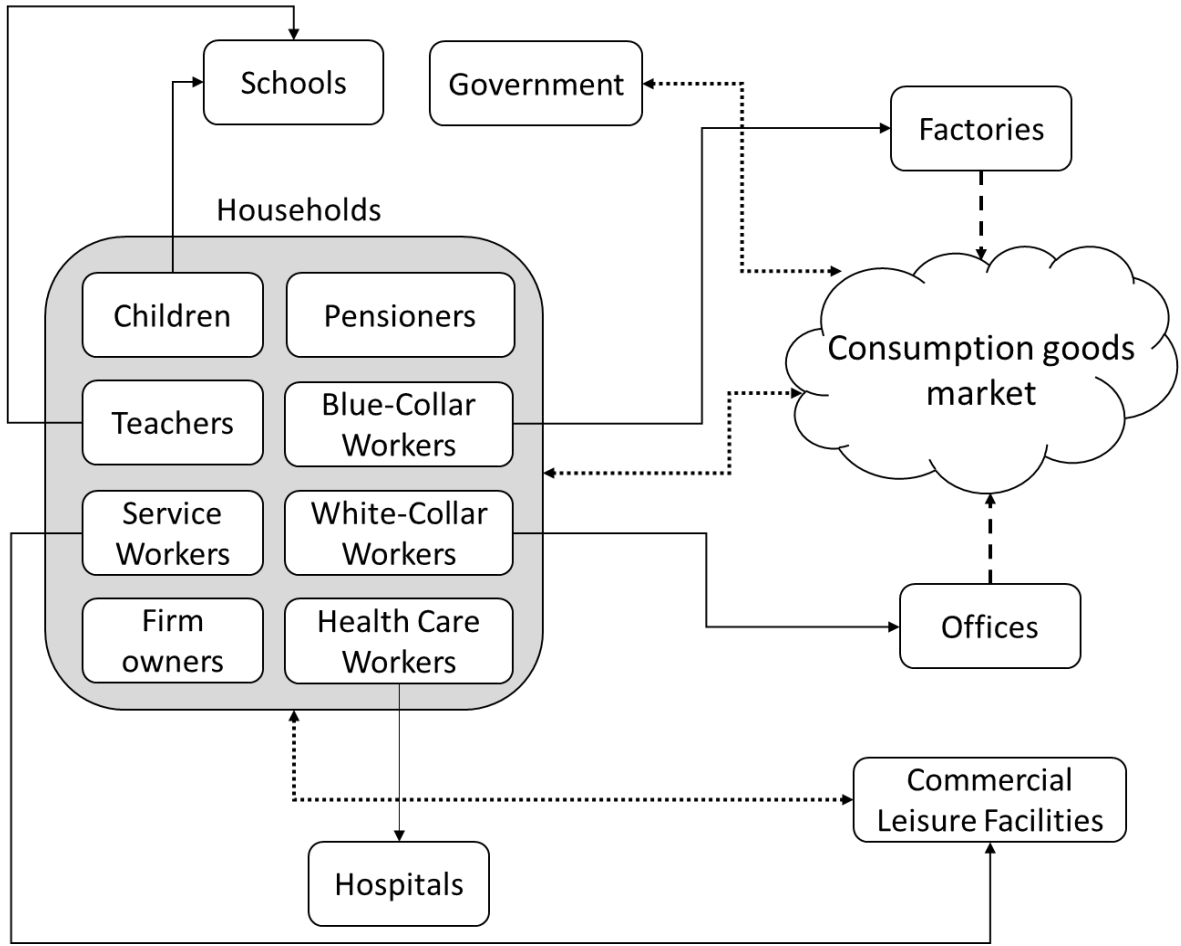

**Figure 4: Depiction of the economy. Solid lines represent work relations, dashed lines production and dotted lines consumption. Factories, offices and commercial leisure facilities are owned by firm owners. Schools and hospitals are operated by the government.**

*2.5.1 Production*

Offices and factories are producing consumption goods using white-collar workers and blue-collar workers respectively.

The production $y_f$ of a factory $f$ is given by the following equation, where $n^I_{f,bc}$ denotes the number of blue-collar workers that are currently present at the factory f and $\alpha_{bc}$ denotes the productivity of blue-collar workers:

$$y_{f,t} = \alpha_{bc} n^I_{f,bc,t}$$

For offices, $n^I_{f,wc}$ denotes the number of white-collar workers that are currently located at the office o. White collar workers also work from home, if they are isolated or in general forced to work from home by government policy. It is assumed that workers who have to take care of little children (i.e. children who belong to the first two age groups) during their work time are less productive than non-care givers and we must thus differentiate between white collar workers, who work from home and do not have to give care $n^{II}_{o,wc,t}$ and those who do have to give care $n^{III}_{o,wc,t}$. Both $n^{II}_{o,wc,t}$ and $n^{III}_{o,wc,t}$ are employed at office o and must be able to work (i.e. not be too sick to work).

The productivity of white-collar workers working at the office is denoted by $\alpha_{wc}$. For those who work from home it is modified with the efficiency parameter $\delta^{II}$, which is between 0 and 1. Those who are care givers experience a further reduction represented by $\delta^{III}$, which is also between 0 and 1.

The production $y_{o,t}$ of an office $o$ is finally given by the following equation:

$$y_{o,t} = \alpha_{wc} \left( n_{o,wc,t}^{I} + n_{o,wc,t}^{II} \delta^{II} + n_{o,wc,t}^{III} \delta^{II} \delta^{III} \right)$$

*2.5.2 Income*

Children (family benefits), all unemployed workers (unemployment benefits), pensioners (pensions), firm owners (rents) and employed white collar workers, blue collar workers and teachers (net wages) are assumed to be paid in phase 1 of each day from Monday until Friday. Employed service and health care workers are paid, whenever they work (by their employer) or should work according to their shift schedule, but are unable to work because they are sick, isolated, their workplaces are closed or they have to give care (by the government).

Unemployment benefits, sick pay, child care benefits and quarantine benefits are fractions of the net wage of a worker of a given type. Gross wages of workers were translated to net wages by calculating the net wages as a fraction of the gross income of a full-time employee (Statistisches Bundesamt 2016). All monetary variables are normalized to the daily gross wage of a service worker.

Formally, $w_{j,t}^{n}$ is equal to the net wage of a certain type of worker, if this worker works in this period and 0 otherwise. The gross income $m_{j,t}^{y}$ of an employed worker, who is able to work, is $m_{j,t}^{y} = w_{j,t}^{n}$. For an unemployed worker it is $m_{j,t}^{y} = \varphi^{II} w_{j,t}^{n}$. For a sick worker claiming sickness benefits it is $m_{j,t}^{y} = \varphi^{III} w_{j,t}^{n}$, for a quarantined worker $m_{j,t}^{y} = \varphi^{IV} w_{j,t}^{n}$ and finally for a care giving worker, who is unable to work due to care giving responsibilities, $\varphi^{V} w_{j,t}^{n}$. Note that white-collar workers, who have to give care, can still work from home and are paid their usual wage rate.

Gross income of firm owners is given as the sum of the rents $r_{i,t}$ paid out by all of their firms $i^{*}$. Since all workers of a given type are assumed to receive a constant, equal wage, their net wages can be set as model parameters. Since income of firm owners is subject to fluctuations, their net income has to be calculated within the model. In order to avoid having to explicitly model the complicated German tax system (which would also in turn make it harder to adapt the model to another country), firm owner income is taxed at a flat 45% rate, which is the German top marginal tax rate. The flat-rate tax is represented by the parameter $\varphi^{I}$.

$$m_{j,t}^{y} = \varphi^{I} \sum_{i^{*}} r_{i,t}$$

The rents paid out by firm $i$ on the other hand depend on the firm's funds $m_{i,t}^{f}$ and the expected rate of profit $\hat{\pi}^{e}$ (more information on $\hat{\pi}^{e}$ is provided in the subsection 2.5.7). The equation is given in a way that it theoretically allows for a circular flow of the economy, in which $m_{i,t}^f$ and $r_{i,t}$ are constant:

$$r_{i,t} = m_{i,t}^f \frac{\hat{\pi}^e}{1+\hat{\pi}^e}$$

Since a week is covering 21 periods, weekly income is defined as $m_{j,t}^{wy} = \sum_{t=t^*-20}^{21} m_{j,t}^y$, where $t^*$ is the current period.

### 2.5.3 Consumption

In the absence of any financial market or investment good, agents want to consume all of their income. In order to smooth consumption over the course of the week, agents reserve 20% of either their weekly income $m_{j,t}^{wy}$ or their current funds $m_{j,t}^f$ – whichever is higher – for consumption $m_{j,t}^c$ on each day from Monday until Friday.

$$m_{j,t}^c = 0.2 * max\ (m_{j,t}^f, m_{j,t}^{wy})$$

Agents are able to consume in two different industries: The consumption goods industry and the leisure industry. Agents choose to save a fraction $h_j$ of $m_{j,t}^c$ for leisure activities, which are added to the savings for leisure facilities $m_{j,t}^l$. To calculate the final value for $m_{j,t}^l$, any expenses for leisure facilities in this period $\widehat{m}_{j,t}^l$ must be subtracted.

$$m_{j,t}^l = m_{j,t-1}^l + m_{j,t}^c h_j - \widehat{m}_{j,t}^l$$

With the rest of the reserved money, agents buy consumption goods.[9] In addition to that, the government uses $g_t$ to buy consumption goods.[10] The market for consumption goods is cleared every day and the price for consumption goods $p_t^c$ is thus given by the aggregated nominal demand divided by the aggregated real supply:

$$p_t^c = \frac{\sum_j (m_{j,t}^c (1-h_j)) + g_t}{\sum_f y_{f,t} + \sum_o y_{o,t}}$$

### 2.5.4 Visits of commercial leisure facilities

Consumption at commercial leisure facilities on the other hand does not behave according to Say's law. Agents only visit them, if they plan to do so and are able to execute their plan, i.e., if they plan to visit a commercial leisure facility or if they plan to visit a friend, who visits a commercial leisure facility. Visits are only possible, if the location is not under quarantine, the agent is not isolated and able to pay for the visit (i.e. $m_{j,t-1}^l + m_{j,t}^c h_j \geq p_{z,t}^l$). If the plan is executed, agents spend $\widehat{m}_{j,t}^l$ at the chosen leisure facility. The expenses depend on the price of the leisure facility $p^l_{z,t}$, but agents also splash out a fraction $\omega$ of their remaining money saved for leisure expenses. The following equation describes $\hat{m}^l_{j,t}$, where $v_{j,z,t}$ denotes the number of visits of leisure facility $z$ (0 or 1). Since agents can only visit up to one leisure facility per period, $\sum_z v_{z,i,t}$ may be 0 or 1.

---

[9] If $m_{j,t-1}^l \geq m_{j,t}^f$, agents do not choose to increase their savings for leisure activities anymore, but use all of $m_{j,t}^c$ to buy consumption goods.
[10] Consumption goods in this model also represent goods which are typically bought by governments such as infrastructure, although this is not modeled explicitly.

$$\hat{m}^l_{j,t} = \sum_z v_{z,i,t}(p^l_{z,t} + \omega(m^l_{j,t-1} + m^c_{j,t}h_j - \sum_z v_{z,i,t}p^l_{z,t}))$$

*2.5.5 Prices of commercial leisure facilities*

Commercial leisure facilities adjust their prices $p^l_{z,t}$ based on their past relative utilization on each Sunday night $t^*$. The rationale behind this is that attractive facilities can increase their prices in order to make better use of their capacities, whereas less attractive facilities decrease their prices. Two kinds of relative utilization are calculated using the absolute utilization $u_{z,t}$, i.e. the number of guests in a given period: $u^I_{z,t}$ denotes the utilization relative to the maximum capacity $\vartheta^I_1$ and $u^{II}_{z,t}$ the utilization relative to the standard capacity $\vartheta^{II}_1$. While the $\vartheta^I_1$ determines whether people are still able to go to this leisure facility (and may be targeted by policies), $\vartheta^{II}_1$ is important for determining the infectiousness of the guests.

$$u_{z,t} = \sum_j \sum_{t=t^*-20}^{21} v_{j,z,t}$$

$$u^I_{z,t} = \frac{u_{z,t}}{\vartheta^I_1}$$

$$u^{II}_{z,t} = \frac{u_{z,t}}{\vartheta^{II}_1}$$

Facilities first consider the utilization relative to the maximum capacity, so if $u^I_{z,t}$ is larger than the parameter $v^I$, $p^l_{z,t} = (1 + \lambda^I)p^l_{z,t-1}$. If considering $u^I_{z,t}$ did not cause any price changes, the firm considers the utilization relative to the standard capacity. If $u^{II}_{z,t}$ is smaller than $v^{II}$, prices are reduced: $p^l_{z,t} = (1 - \lambda^{II})p^l_{z,t-1}$. If $u^{II}_{z,t}$ is higher than $(1 - v^{II})$, prices are increased: $p^l_{z,t} = (1 + \lambda^{II})p^l_{z,t-1}$.

*2.5.6 Profits*

Profits are revenues minus costs and are computed on a weekly basis on each Sunday night. The government is assumed to replace (fractions of the) net wages of those workers who are unable to work because they are sick $n^{IV}_{f,bc,t}$, quarantined $n^V_{f,bc,t}$ or have to take care of a child $n^{VI}_{f,bc,t}$.[11] The parameters $\varphi^{III}$, $\varphi^{IV}$ and $\varphi^V$ describe, which fraction of the wage is replaced respectively. Revenues $s_{o/f/z,t}$ are thus assumed to equal sales plus wage replacements.

For offices and factories, sales are price times produced quantity.

---

[11] Please note that $n^{VI}_{f,bc,t} = 0$ for offices, because care taking service workers are assumed to still work from home – albeit with less productivity.

$$s_{f,t} = y_{f,t} p_t^c + w_{bc,t}^n (n_{f,bc,t}^{IV} \varphi^{III} + n_{f,bc,t}^{V} \varphi^{IV} + n_{f,bc,t}^{VI} \varphi^{V})$$

$$s_{o,t} = y_{o,t} p_t^c + w_{wc,t}^n (n_{o,wc,t}^{IV} \varphi^{III} + n_{o,wc,t}^{V} \varphi^{IV})$$

For commercial leisure facilities, sales are equal to the sum of the expenses of agents who visited this leisure facility in a given period.

$$s_{z,t} = \sum_j v_{j,z,t} \widehat{m}_{j,t}^l + w_{sw,t}^n (n_{z,sw,t}^{IV} \varphi^{III} + n_{z,sw,t}^{V} \varphi^{IV} + n_{z,sw,t}^{VI} \varphi^{V})$$

In this simple model, costs only consist of labor costs, which are gross wages of workers $w_{wc/bc/sw}^g$, multiplied with the number of employees who were scheduled to work for the firm in the current period $n_{o/f/z,wc/bc/sw,t}$.

Costs for factories f, offices o and commercial leisure facilities z are thus given as:

$$c_{f,t} = w_{bc}^g n_{f,bc,t}^I + w_{bc,t}^n (n_{f,bc,t}^{IV} \varphi^{III} + n_{f,bc,t}^{V} \varphi^{IV} + n_{f,bc,t}^{VI} \varphi^{V})$$

$$c_{o,t} = w_{wc}^g (n_{o,wc,t}^I + n_{o,wc,t}^{II} + n_{o,wc,t}^{III}) + w_{wc,t}^n (n_{o,wc,t}^{IV} \varphi^{III} + n_{o,wc,t}^{V} \varphi^{IV})$$

$$c_{z,t} = w_{sw}^g n_{z,sw,t}^I + w_{sw,t}^n (n_{z,sw,t}^{IV} \varphi^{III} + n_{z,sw,t}^{V} \varphi^{IV} + n_{z,sw,t}^{VI} \varphi^{V})$$

Finally, weekly profits for firm i $\pi_{i,t}$ and its rate of profit $\widehat{\pi}_{i,t}$ are computed on each Sunday night $t^*$ as:

$$\pi_{i,t} = \sum_{t=t^*-20}^{21} (s_{i,t} - c_{i,t})$$

$$\widehat{\pi}_{i,t} = \frac{\pi_{i,t}}{\sum_{t=t^*-20}^{21} c_{i,t}}$$

### 2.5.7 Labor market

Firms (factories, offices, commercial leisure facilities) decide whether they want to hire or fire workers based on the difference between their actual $\widehat{\pi}_{i,t}$ and expected rate $\widehat{\pi}^e$ of profit and a buffer parameter $\varepsilon$. If $\widehat{\pi}_{i,t} - \widehat{\pi}^e > \varepsilon$, the firm tries to hire one worker, if $\widehat{\pi}^e - \widehat{\pi}_{i,t} > \varepsilon$, the firm will fire one worker.[12] If a firm only has a single worker left, it will only choose to fire (and thus end the existence of the firm), only if the firm's funds are negative. Workers always accept job offers, i.e. firms are always able to fill their job offers as long as there are still enough unemployed workers of a given type left.

### 2.5.8 Government

---

[12] This formulation obviously only makes sense due to the small firm sizes. If firms were larger / more heterogeneous, it would make sense to introduce a parameter that indicates how many workers as fractions of the current staff should be hired/fired.

The government gains money from the income tax and pays for pensions, family benefits, unemployment benefits, quarantine benefits and the wages of health care workers and teachers. Additionally, the government uses $g_t$ to buy consumption goods.

Since the government only consumes 5 times a week, but has revenues in 14 periods of the week and expenditures in 21 periods per week (because it pays salaries of health care workers), the equations describing the behavior of the government must include a variable for government savings $g_t^S$, which tracks its revenues and expenses.

The following equation describes $g_t^S$, where $q^I$ comprises the employed workers in the private sector (white collar workers, blue collar workers, service workers), $q^{II}$ the firm owners, $q^{III}$ the workers in the government sector (health care workers, teachers), as well as children and pensioners, $q^{IV}$ the unemployed, $q^V$ the sick employees, who claim sickness benefits, $q^{VI}$ the quarantined, who gain quarantine benefits and finally $q^{VII}$ care givers, who receive care giving benefits.

$$g_t^S = g_{t-1}^S - g_{t-1} + \sum_{q^I}(w_{q^I,t}^g - w_{q^I,t}^n) + \sum_{q^{II}} \varphi^I m_{q^{II},t}^y$$
$$- \sum_{q^{III}} w_{q^{III},t}^n - \sum_{q^{IV}} \varphi^{II} w_{q^{IV},t}^n - \sum_{q^V} \varphi^{III} w_{q^V,t}^n - \sum_{q^{VI}} \varphi^{IV} w_{q^{VI},t}^n - \sum_{q^{VII}} \varphi^V w_{q^{VII},t}^n$$

I test two stylized government spending policies:

*2.5.8.1 Zero-deficit*

In this case, the government aims for a zero deficit, i.e. sets government purchase of consumption goods $g_t$ such that it equals $g_t^S$ (as long as $g_t^S \geq 0$). This policy acts procyclically, since government purchases are increased if more people are employed and working.

$$g_t = \max(0, g_t^S)$$

*2.5.8.2 Fixed government purchase*

Here, the government always uses $g_0$ to purchase consumption goods. This policy acts anticyclical, because the government borrows money, if a lot of people are out of work and saves money in an upswing. $g_0$ is calculated to match a circular flow of the economy in the initialization period (see the next chapter for details).

$$g_t = g_0$$

**3 Model set up and calibration**

I parametrized the model such that the parameters at least roughly fit their empirical counterparts quantitatively or – where parameters cannot be obtained, calculated or estimated due to the lack of data – qualitatively, i.e. in a stylized way. As time passes, more and more data about the COVID-pandemic is published, which means that the parametrization can be improved in the future.

I use statistical data to generate a virtual town – COVID-Town, which aims to represent a smaller version of Germany at the scale 1:1000. Since this process, as well as the simulation itself, involves a large number of stochastic processes, I repeat the simulations a large number of times (a process known as Monte Carlo simulations) in order to avoid putting too much emphasis on single runs, which may be outliers. These simulations are carried out with fixed random seeds, which has the advantage that the same town can be exposed to alternating policy scenarios, which gives the simulation approach an edge over empirical investigations, which are limited by a) actually existing policy heterogeneity and b) possibly high country-specific heterogeneity, which influences the results.

The medical characteristics are taken from the LMIC reports of the Imperial College (2020), which relies mostly on UK data. Exceptions are made for a) the duration of a mild case, which is assumed to be one week (compared to a very low number of 1.6 days, which I found nowhere else in the literature and for which the authors do not provide any source), b) a difference was made between the latent and the incubation period, since this is easily possible in an ABM and allows the model to capture an additional stylized fact (presymptomatic infections) and c) the probability of a critical case to die without receiving intensive care is assumed to be 100% instead of 95%. The age-specific medical parameters were already given in table 2.

**Table 3: Medical parameters**

|   |   |
|---|---|
| Incubation Period | 15 |
| Latent period | 13 |
| Duration of a mild case | 21 |
| Duration of a critical or severe case until hospital admission | 12 |
| Duration of a severe case, if the result is surviving | 29 |
| Duration of a severe case, if the result is dying | 23 |
| Duration of a critical case, if the result is surviving | 34 |
| Duration of a critical case, if the result is dying | 30 |
| Duration of a formerly critical case, which is recovering from ICU | 10 |
| Probability of dying in an ICU | 50% |
| Probability of a critical case dying without an ICU | 100% |
| Probability of a severe case dying without a hospital bed | 60% |

**Source**: Imperial College (2020), assumptions (duration of a mild case, probability of a critical case dying without an ICU), Time in phases (3 phases = 1 day)

The challenge of calibrating the epidemiological parameters of the model is that data on the real number of infections per day instead of the number of reported infections per day is not available. In order to estimate the former, one has to overcome two information problems:

1.) Reported infection data is lagging, because infected are typically only tested after developing symptoms and the analysis as well as reporting of the results cause further delay. The "nowcasting" by the Robert Koch Institut (2020a, 2020b) accounts for these difficulties and dates back the infection date of each positively tested infected to an estimated true infection date by means of multiple imputation.

2.) Not every infected is tested, either because a) they are asymptomatic and do not suspect to be infected, or b) because the testing capacities are exhausted, or c) because they consciously decide not to test themselves. If we assume, however, that the number of reported deaths (by Robert Koch institute 2020c) approximately equals the true number of COVID-19 related deaths, we can estimate the dark figure of infected by running simulations, in which the number of simulated deaths equals the empirically observed number of deaths. Assuming that the number of observed deaths equals the actual number of COVID-related deaths seems to be justified regarding the German case for two reasons: a) severe and critical COVID-19 cases require hospital care, where they are likely tested. Hospital capacities were not overloaded in Germany and universal health care is in place, so we can assume that everybody who needed hospital care actually got it and b) excess mortality[13] in Germany is very close to the reported number of COVID 19 deaths (see Statistisches Bundesamt 2020b).

If we assume that the dark figure captured by 2.) is constant over time, i.e. the detection rate is constant, we can estimate it by calibrating the model to a dark figure that produces the observed deaths. In order to ensure an internally consistently validated input, the detection rate in the model has to be equal to the empirically estimated one. This means that a detection rate of 50% implies that the dark figure is as high as the figure reported by Robert Koch Institut (2020b). This condition makes calibration particularly challenging, since all epidemiological parameters must be calibrated at the same time.

The following parameters were calibrated to fit the empirical data on infections (first 50 days) and deaths (first 100 days) quantitatively and the relative infection risks qualitatively:

**Table 4: Epidemiological parameters**

| Description | Value |
|---|---|
| Infected at period 0 | 0.007% |
| Baseline infection probability $\beta$ | 9.5% |
| Maximum social contacts $\gamma$ | 10 |
| Detection threshold | 66.6% |
| Unable to work threshold | 70% |
| Hygienic parameter $\pi_p$ | 1 |
| Standard capacity of commercial leisure facilities $\vartheta_1^{II}$ | 8 |
| Standard capacity of non-commercial leisure facilities $\vartheta_2^{II}$ | 800 |
| maximum capacity | 400% |

**Explanation**: The default capacity of commercial leisure facilities is calculated in the following way: the expected number of guests are divided by the number of main leisure shifts (9) + the number of off-peak leisure shifts divided by 2 (2.5). The sum is then multiplied by 0.5 (because it is assumed that infection is more likely in these locations). Up to 400% (i.e. 32 for a commercial leisure facility) guests fit into a leisure facility at a given period. A sensitivity analysis for the baseline infection probability is shown in appendix A.

---

[13] The difference between total deaths in 2020 and the average total deaths in the last couple of years (Statistisches Bundesamt 2020b uses the average of 2016-2019).

The economic parameters were largely calibrated to match their empirical counterparts. The expected profit rate is – in absence of capital goods in this model – calculated as operating surplus divided by gross wages. The buffer parameter and the price adjustment parameters were set so as not to produce drastic changes in absence of the virus. The teleworking parameters were set in a way that offices are not disadvantaged compared to factories (which would contradict the empirics):

**Table 5: Economic parameters**

| Description | Value | Source |
|---|---|---|
| Expected profit rate $\hat{\pi}^e$ | 0.4 | Statistisches Bundesamt (2019c), p. 334 |
| Profit rate buffer ε | 0.1 | Assumption |
| Unemployment benefits $\varphi^{II}$ | 0.6 | Bundesministerium für Gesundheit (2020) |
| Sick pay $\varphi^{III}$ | 1 | Bundesministerium für Gesundheit (2020) |
| Quarantine benefits $\varphi^{IV}$ | 1 | Bundesministerium für Gesundheit (2020) |
| Caregiving pay $\varphi^V$ | 0.67 | Bundesministerium für Gesundheit (2020) |
| Teleworking efficiency | 1 | Assumption |
| Care giving teleworking efficiency | 0.8 | Assumption |
| Price adjustment parameter 1 $\lambda^I$ | 0.05 | Assumption |
| Price adjustment parameter 2 $\lambda^{II}$ | 0.02 | Assumption |
| Leisure money splash parameter $\omega$ | 0.4 | Assumption |
| Productivity of blue-collar workers $\alpha_{bc}$ | 1 | Normalized |
| Productivity of white-collar workers $\alpha_{wc}$ | $\frac{1.77}{1.28}$ | Assumption: no wage discrimination |

A sensitivity analysis for the profit rate buffer is shown in appendix B.

In the beginning of the model set up, one location is created for each household, office, factory, school, hospital, retirement home, commercial and non-commercial leisure facility. Then, the non-household locations are filled. Their numbers are mostly calculated from publicly available statistical data:

**Table 6: Number of locations and their characteristics**

| Description | Value | Source |
|---|---|---|
| Number of employed blue-collar workers per factory: | 12 | Statistisches Bundesamt (2020a) |
| Number of employed white-collar workers per office: | 10 | Statistisches Bundesamt (2020a) |
| Number of employed service workers per commercial leisure facility: | 4 | Statistisches Bundesamt (2020a) |
| Number of teachers per school | 32 | Statistisches Bundesamt (2020a) |
| Number of hospitals per capita | 1942 / 82158111 | Statistisches Bundesamt (2018a) |

| Number of retirement homes per capita | 14480 / 82158111 | Statistisches Bundesamt (2018b) |
|---|---|---|
| Average school class size | 22 | OECD (2020a) |
| Number of non-commercial leisure facilities per commercial leisure facility | 1 | Assumption |
| Number of hospital beds per 1k inhabitants | 8 | OECD (2020b) |
| Number of ICUs per 100k inhabitants | 36.6 | DIVI (2020) |

In the next step, the households described in the following table are created:

**Table 7: Number and types of households**

| Description | Value |
|---|---|
| Number of households per capita | 0.5 |
| single households | 13.8% |
| singles with kids | 6.2% |
| couples without kids | 17.3% |
| couples with kids | 29.2% |
| intergenerational household without children: | 4.2% |
| intergen household w children | 1.5% |
| single pensioners | 15.9% |
| couples of pensioners: | 11.9% |

**Source**: Statistisches Bundesamt (2019b)

Then, the households and retirement homes are populated with 82000 human agents – each agent thus represents one thousand real German inhabitants. Their age distribution was already given in table 2. Employed workers are assigned to workplaces. The share of each type, the unemployment rate and wages are given in table 8:

**Table 8: Share of agents, unemployment, wages**

| Description | Profession share | Initial unemployment | Gross wage | Net wage |
|---|---|---|---|---|
| Children/young people | 17.57% | - | 0.08 | 0.08 |
| Blue-collar workers | 20.24% | 10.7% | 1.28 | 0.81 |
| White-collar workers | 23.96% | 5.5% | 1.77 | 1.05 |
| Service workers | 2.69% | 17.4% | 1 | 0.66 |
| Teachers | 6.62% | 8.8% | 1.39 | 0.86 |
| Health care workers | 6.42% | 3% | 1.49 | 0.91 |
| Pensioners | 21.5% | - | 0.32 | 0.32 |
| Firm owners | 1% | - | - | - |

**Sources**: The top 1% are assumed to be firm owners. Share of children and pensioners from Statistisches Bundesamt (2019a). Shares of workers and initial unemployment calculated from Bundesagentur für Arbeit (2020a) and Bundesagentur für Arbeit (2020b). Gross wages from Statistisches Bundesamt (2016). Gross wages are normalized to the wage of a service worker. Net wages are calculated using an online calculator: http://www.parmentier.de/steuer/gehaltsrechner14.htm. Children are assumed to receive child care benefits. Average pension from Bundesministerium für Arbeit und Soziales (2015). Wage data from 2014 is necessary, since Statistisches Bundesamt (2016) is the most recent publication showing wages for jobs according to the "Klassifikation der Berufe 2010" which was used to determine the shares of the worker types.

For pensioners, additional information about their residence is necessary, because they may also live in retirement homes:

**Table 9: Residence of pensioners**

| Description | Value | Source |
|---|---|---|
| Pensioners in intergenerational households | 15.3% | Statistisches Bundesamt (2019a, 2019b) |
| Pensioners in retirement homes | 4.5% | Statistisches Bundesamt (2018b, 2019a) |
| Pensioners in pensioner-only households | 80.2% | Remaining from 100% |

**Sources**: Pensioners in intergenerational households were calculated by dividing the difference between total number of people above 65 and people living only with cohabitors of 65+ from Statistisches Bundesamt (2019b) by the total number of people above 65 from Statistisches Bundesamt (2019a). Please note that this number includes households that do not fit the conventional definition of an intergenerational household, e.g. a married couple in which one spouse is above 65 and the other isn't. This definition is necessary, however, to remain consistency with the types of households described in table 7. Pensioners in retirement homes were calculated by dividing the number of people who require inpatient care at retirement homes by the total number of people above 65.

Initial income of firm owners is calculated as the average income of firm owners, which is the sum of the gross wages of all workers in the private sector times the expected rate of profit divided by the number of firm owners. It is thus not calibrated to match real data, but to match a circular flow of the economy.

Then, the leisure preferences for the human agents are set and edges to leisure facilities and friends (other agents) are created and weighted. According to a survey by Sinus Institut (2018), Germans have on average 3.7 close friends. As an approximation, I assume that the number of friends is described by a uniform distribution from 1 to 6. This number is drawn for all agents and afterwards edges are randomly created to match the number of friends drawn for each individual. For simplicity and due to the lack of data, I assume that the number of edges to non-commercial leisure facilities has the same mean and is either 3 or 4 for each individual. The age-specific preferences for each leisure activity are then derived from the time spent on each activity following the German time use study (Statistisches Bundesamt 2015). Since the friendships described by the edges between human agents can connect agents who do not belong to the same age group, their weight is normalized to 50 and all other preferences are calculated accordingly (i.e. the sum of all weights is higher for age groups, which have a relatively low preference for meeting friends).

**Table 10: Leisure preferences**

| | Mean μ | |
|---|---|---|

| Age group | Social contacts | non-commercial leisure facilities | commercial leisure facilities | Staying home | standard deviation σ | E (Sum) |
|---|---|---|---|---|---|---|
| 2-3 | 50 | 42 | 48 | 396 |  | 826 |
| 4-5 | 50 | 36 | 76 | 450 |  | 922 |
| 6-8 | 50 | 46 | 72 | 582 | 10% | 1080 |
| 9-12 | 50 | 56 | 60 | 679 |  | 1185 |
| 13-16 | 50 | 68 | 64 | 810 |  | 1367 |

**Sources**: Calculated using the German time use study (Statistisches Bundesamt 2015). The expected sum is calculated multiplying the mean of social contacts, non-commercial leisure facilities and commercial leisure facilities (3.5) with the average weight and summing up the preferences. The means for leisure facilities are furthermore divided by the mean attractiveness parameter $\mu^l$ (not shown in this table in order to make these columns comparable to the others), which is assumed to be 5. The utility gained from commercial leisure facilities includes the multiplier $\kappa$, which is assumed to be 2 and which accounts for the idea that the observed preference for going e.g. to a restaurant is lower than the actual one, because of budgetary restrictions. The standard deviation is assumed to be 10% of the respective mean, i.e. e.g. 5 for social contacts. Age groups 0-1 (i.e. 0-9 year olds) are not included in the time use study. They are thus assumed to spend their leisure with a chaperone from the same household, who is at least 10 years old and do not save any money for leisure activities. A sensitivity analysis is shown in appendix C.

Prices of leisure facilities and the fraction saved by each individual for leisure activities are calculated to match approximately a circular flow of the economy using the expected visits, number of employees for each firm and the expected profit rate $\hat{\pi}^e$.

In order to do this, the program first calculates an expected weekly revenue per commercial leisure facility, which is equal to the expected weekly wage costs (i.e. five times the number of employees per commercial leisure facility times the daily gross wage of service workers) times the expected profit rate. The default price is then calculated as the expected weekly revenue of the whole leisure industry divided by the expected weekly visits of the whole population. The fraction saved by each individual for leisure activities is then the expected weekly visits of individuals of the same age group times the default leisure facility price divided by the average income of the profession (including unemployed). For firm owners, the fraction is assumed to be twice as high: both because they have more leisure periods than employed workers and because they are assumed to spend more money during each visit.

The funds of factories and offices are set to cover the wage expenses of one working day. The funds of commercial leisure facilities are set to $(2 + \hat{\pi}^e)n_{z,sw,t}w_{sw}^g$, where $n_{z,sw,t}$ denotes the number of service workers employed at the commercial leisure facility and $w_{sw}^g$ their gross wage rate.[14]

All workers go to work, produce consumption goods (if applicable) and get paid. Unemployment benefits are paid.

---

[14] This difference is necessary, because during the setup phase, wages are paid and consumption goods are produced and consumed in order to establish a baseline level of production, but agents do not visit leisure facilities.

Finally, the initial government savings = expenditure is set to match the expected revenues of offices and factories (i.e. their wage costs times the expected rate of profit) minus the expected private consumption in order to ensure a circular flow in the setup period (t=0):

$$g_0 = \left(\sum_f c_{f,0} + \sum_o c_{o,0}\right)\hat{\pi}^e - \sum_j m^c_{j,0}$$

The consumption goods market is cleared, firm owners receive their rents and all agents go home.

**4 Policies**

The model allows for high flexibility regarding the explicit implementation of policies. For a start, I implemented stylized versions of actually used policies to create a baseline scenario, which was used to calibrate the model.

*Increased sanitary standards at hospitals*

This policy lowers the hygienic parameter at hospitals to 0.1, i.e. the infection probability is only one tenth of the probability of becoming infected at home.

*Isolation*

Every infected with a severity level above a detection threshold is isolated from onset of the symptoms until the end of the infection.

*Family isolation*

If a symptomatic agent is detected, its household members are isolated for a given number of periods. Following the German policy of isolating for 2 weeks, the isolation duration is set to 42 periods.

*Workplace isolation*

Works like the policy on family isolation, but targets the co-workers of an agent. Workplace isolation only applies to white collar workers, blue collar workers and service workers.

*School closures*

All schools are closed down and children, as well as teachers, do not move to school locations anymore during their work phases. Children under the age of 10 require a care giver to stay at home with them.

*Commercial leisure facility closures*

All commercial leisure facilities are closed down. Agents may still plan to visit them during their leisure phase, but these plans will fail. Service workers will stay at home during their work phases.

*Social distancing*

This policy accounts for the fact that agents voluntarily engage in social distancing to avoid infections.

In the scenarios presented in this paper, this happens via two ways:

1.) The home preference multiplier described in subsection 2.3 is increased to 2, i.e. agents prefer to stay at home during their leisure phases more often.
2.) They reduce their maximum number of social contacts per period. SORA (2020) conducted a survey in Austria, which showed that the participants who worked from their workplaces during the lockdown measures spent 44% less time in close contact with people not living in their households. Only 17% of this group reported to have 1-5 daily contacts before, but 65% after the introduction of the policies. German data on this topic is currently not available, but I assume a similar effect. As an approximation, the maximum number of social contacts is reduced by 50%.

*Contact ban*

Agents may not meet their friends anymore.

*Teleworking mandatory*

White collar workers start to work from home.

## 5 Verification

Model verification is the process of exploring how a model works and whether it works in the way that it is supposed to (Gräbner 2018). This model was created in a stepwise fashion, i.e. in the beginning, agents were only located at their households and could interact at home, then they were able to meet their friends etc., and at the end an increasingly complex economy was introduced. This step wise creation helped to verify the model by testing the added functions individually to eliminate programming errors.

The simple specification of the economic part of the model was in addition to that used to verify the model with regard to three aspects:

1.) The funds of all agents, firms and the government plus the funds saved for leisure activities must remain constant throughout the simulation (stock-flow consistency).
2.) In absence of the virus, the funds of the agents should only change little with respect to the initial setup. The remaining changes should be traced back to a) the different timing of wage payments and b) the unstable demand in the leisure industry (circular flow).
3.) The rate of unemployment in the leisure industry should – in absence of the virus – hover around the initial rate of unemployment. Changes in the beginning of the simulation should be more drastic than after a while, since leisure facilities start out with a homogenous labor force, but heterogeneous demand (due to a different attractiveness factor and different links to human agents) and adapt their labor force and prices to their demand after a while (circular flow).

**6 Validation**

Model validation tests whether the model, its inputs and its outputs correspond to the real world (Rand and Rust 2011, Tesfatsion 2017, Gräbner 2018). Tesfatsion (2017) considers four ways to empirically validate a model: input validation, process validation, descriptive output validation and predictive output validation. I used all four approaches to test the model's capacity to insert, include, produce and predict stylized facts. The list of stylized facts replicated by this model presented in this paper is necessarily incomplete, as not every empirical stylized fact is described in the literature yet or known to this author. COVID-Town can be used in the future, however, to check whether it is able to replicate other stylized facts.

In addition to the validation with the use of stylized facts, I use (limited) quantitative input and descriptive output validation, i.e. the model is calibrated with a low number of assumed / manually calibrated parameters and the simulation results fit to the observed German death curve and estimated infection curve the of the first wave quantitatively.

*6.1 Input validation*

The goal of input validation is to make sure that the inputs of the model are empirically relevant. During the stepwise creation of the model (as described in *verification*), more and more inputs that seemed to be highly important to the epidemiological model were added. The final model includes at least the following stylized facts:

1.) There are multiple possible sources of infection, which are modeled explicitly: leisure, home, retirement home, work, hospital, school.[15]

2.) Agents differ with regard to their age and the risk of a severe or critical course increases with it (see Verity et al. 2020).

3.) Agents differ with regard to their profession / social class, which determines their social contacts and ability to engage in social distancing.

4.) There is a difference between latent and incubation period, which means that the model can capture presymptomatic infections.

5.) Severe cases require hospital beds and critical cases an ICU. Both may result in death, although the probability is much higher if the capacity of hospital beds/ICUs is surpassed.

6.) There are several types of households. Pensioners living in intergenerational households are less able to engage in social distancing than pensioners without coresidents who are working.

7.) There are industries, in which workers can work from home. In other industries, workers must work on-site. In some of the latter, workers must interact with customers, which pose an additional risk of infection.

---

[15] My model includes all but one (traveling) sources of infections listed by the Austrian Agency for Health and Food Safety in their cluster analysis. As of the 17th of August 2020, the agency was able to trace back 11,236 of 23,516 confirmed COVID cases to one or more sources of infection. Travelling accounted for 465 cases (see Agentur für Ernährungssicherheit 2020).

8.) Young children require care, if schools/child care facilities close.

9.) Caring parents cannot go to work and are less productive if they are able to work from home.

10.) Agents have varying preferences on how to spend their leisure. The relevant leisure contacts are most often not completely random, but conscious decisions to meet friends or go to specific leisure facilities like bars and clubs based on preferences.

11.) The risk of becoming infected in a crowded location is higher than in a location that permits visitors to keep their distance.

12.) The infection risk at open-air locations (represented by non-commercial leisure facilities) is much lower than at indoor locations.

13.) Agents engage in social distancing via two ways: they reduce their time spent at locations, where they could meet other agents (as seen e.g. in the google mobility reports) and when they go to such places, they reduce their contacts with other agents (as shown by SORA 2020 for the Austrian case).

Furthermore, I parametrized the model – as far as possible – with German statistical data to create a model town that represents Germany quantitatively with regard to the characteristics described in the section on calibration.

*6.2 Process validation*

Process validation considers, whether the processes implemented in the model reflect real-world processes (Tesfatsion 2017). The model incorporates at least the following stylized facts:

1.) Agents meet and may become infected in explicitly defined locations.

2.) Agents spend time in these locations based on their profession and the current day and time.

3.) The infection risk depends on whether protective measures are taken and whether a location is crowded or not.

4.) The model departs from the homogenous mixing assumption by introducing heterogeneous mixing: The chance of an agent to meet another is determined by their profession, leisure preferences and possible friendships (with household members).

5.) Containment policies are modeled explicitly. For instance, if the government decides to close schools, children and teachers will stay at home. This likely has a different effect than to simply assume that this policy reduces social contacts by X%.

6.) The model is stock-flow consistent. In the absence of any money creation, the sum of the funds of human agents, firms and the government remains constant throughout the simulation.

*6.3 Descriptive output validation*

Descriptive output validation analyzes how well a model can be calibrated to fit a certain dataset (Tesfatsion 2017). The model at least replicates the following stylized facts:

1.) Health care workers are exposed to a high risk of infection at work, since their work involves close contact with infected, especially if the hygienic measures at hospitals are not very strict.

2.) The COVID crisis has an impact both on supply and demand of the economy (see section 7 for more details).

3.) Outbreaks in retirement homes are especially dangerous, since they involve the most vulnerable age groups.

In addition to these stylized facts, the model is able to fit the estimated empirical curves of new infections and total deaths in Germany (see fig. 5 and 6). As explained in the calibration section, I calibrated the model to fit a) the empirically observed deaths curve and b) the RKI nowcasting daily infections (Robert Koch Institut 2020b) multiplied with a factor x accounting for the dark figure. Finding a factor, which fits to both curves gives the model's estimation of the dark figure. In order to ensure internal consistency, the factor x has to be equal to one divided by the detection rate. Based on this approach, I estimate the dark figure to be around 2, i.e. for each detected infected, there are two infections which remain undetected. These simulation results are based on the baseline scenario presented in the next section, which aims to mimic the empirical German policy response in a stylized way.

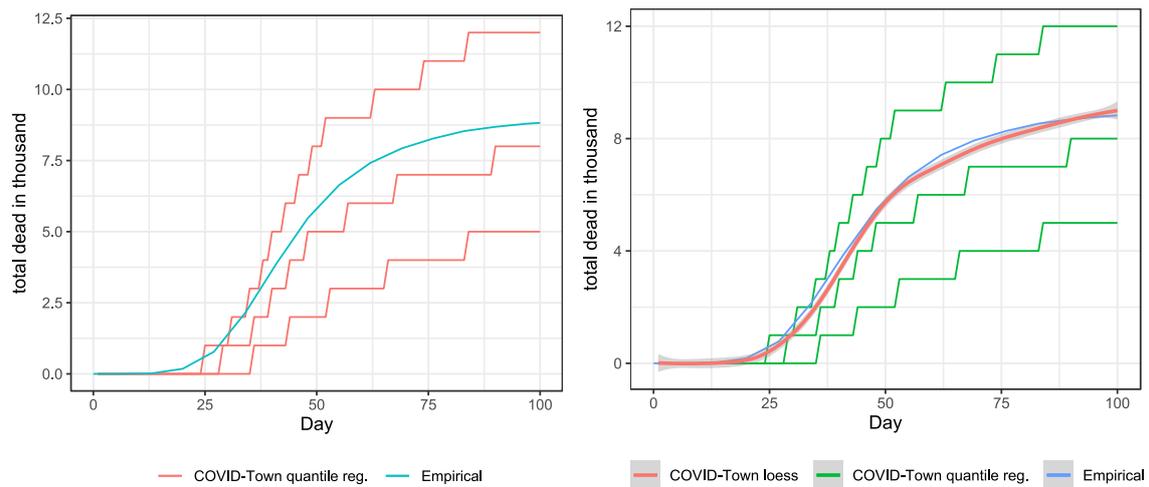

**Figure 5: Simulation quantile regression results and empirically observed deaths according to Robert Koch Institut (2020c) (left), quantile regression, loess and empirics (right)**

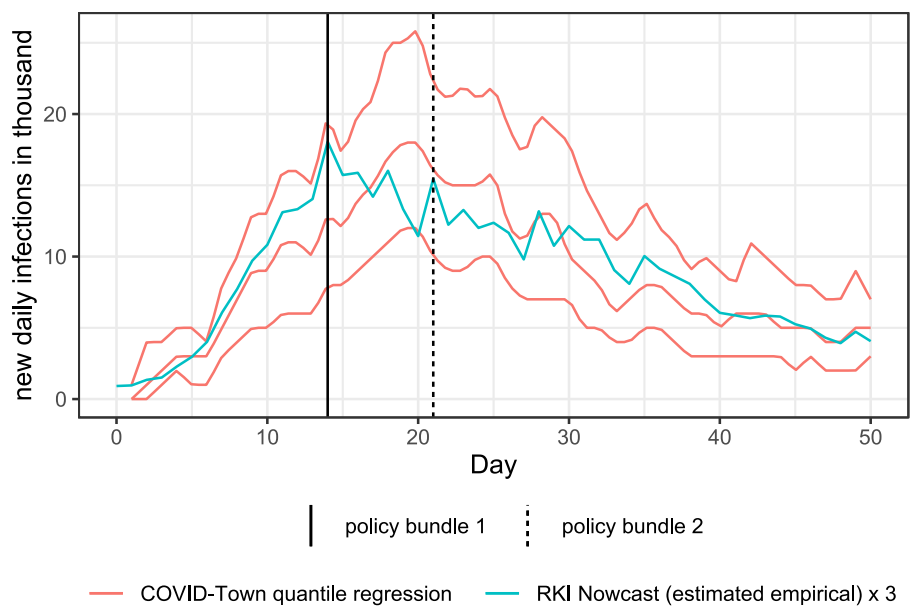

**Figure 6: Simulation quantile regression results and estimated dark figure-adjusted daily infections by Robert Koch Institut (2020b)**

*6.4 Predictive output validation*

Predictive output validation considers, how well the output of a model can forecast data that was not used to calibrate it (Tesfatsion 2017). An obvious route to action would be to calibrate the model with data from another country to see, whether it is also able to predict the observed number of deaths for this country. Due to the time and space necessary to collect this data and describe it, this approach is outside of the scope of this paper. Instead, I validate the predictive output of the model by checking whether the counterfactual policy scenarios described in the next section generate output that matches empirical relations in a stylized way, as a quantitative replication is impossible by definition. The counterfactual policy scenarios produce at least the following two stylized facts:

1.) Early implementation of can reduce the number of infections and deaths (Huber and Langen 2020).

2.) Expansionary fiscal policy can constrain the economic losses (Deb et al. 2020).

**7 Results**

In this section, I first present the results of three containment scenarios to then move on to two stylized fiscal policy scenarios. The baseline scenario was designed to mimic the empirical German policy response, albeit in a simplified manner. It was necessary to simplify, since a) many German policies against COVID-19 such as school closures were not taken at a federal, but at a regional or even local level and b) the model includes by far less details than it would need to fully represent the actual policies. This is even more true for the relaxation of the containment policies, which is why I calibrated the scenario to fit the first 50 days (until the first major policy relaxations).

After calibrating the model to the baseline scenario, I test two alternative containment scenarios: In the rapid action scenario, all policies are implemented one week earlier, in the delayed action scenario one week later. All scenarios start at the evening of the 2$^{nd}$ of March 2020 (Monday), which is the setup period.

Each scenario in this section is simulated 500 times with different random seeds. Since each person in the model represents 1000 people in real life, it is possible that the virus is eliminated in the model, where it wouldn't be eliminated in the real world. In order to avoid biases stemming from this characteristic of the model, I only analyze those simulation runs, in which at least one person is infected at period 300, i.e. at the end of day 100.

**Table 11: Containment policy scenarios**

|  | **Baseline scenario** | **Rapid action** | **Delayed Action** |
|---|---|---|---|
| 02.03.2020 (day 0) | Increased sanitary standards at hospitals, isolation, family isolation, workplace isolation | | |
| 09.03.2020 (day 7) | - | schools, comm. leisure fac., SD | - |
| 16.03.2020 (day 14) | schools, comm. leisure fac., SD | contact ban, teleworking mandatory | - |
| 23.03.2020 (day 21) | contact ban, teleworking mandatory | - | schools, comm. leisure fac., SD |
| 30.03.2020 (day 28) | - | - | contact ban, teleworking mandatory |

**Sources:** Bundesregierung (2020a, 2020b), Welt (2020), **Description**: schools = schools closure, comm. Leisure fac. = commercial leisure facilities close, SD = social distancing

After calibrating the model, I test two alternative containment scenarios: In the rapid action scenario, all policies are implemented one week earlier, in the delayed action scenario one week later. All scenarios start at the evening of the 2$^{nd}$ of March 2020 (Monday), which is the setup period.

Figure 7 and 8 show quantile regression results of the Monte Carlo simulations compared to the estimated empirical infection curve, as well as to the observed empirical deaths curve.

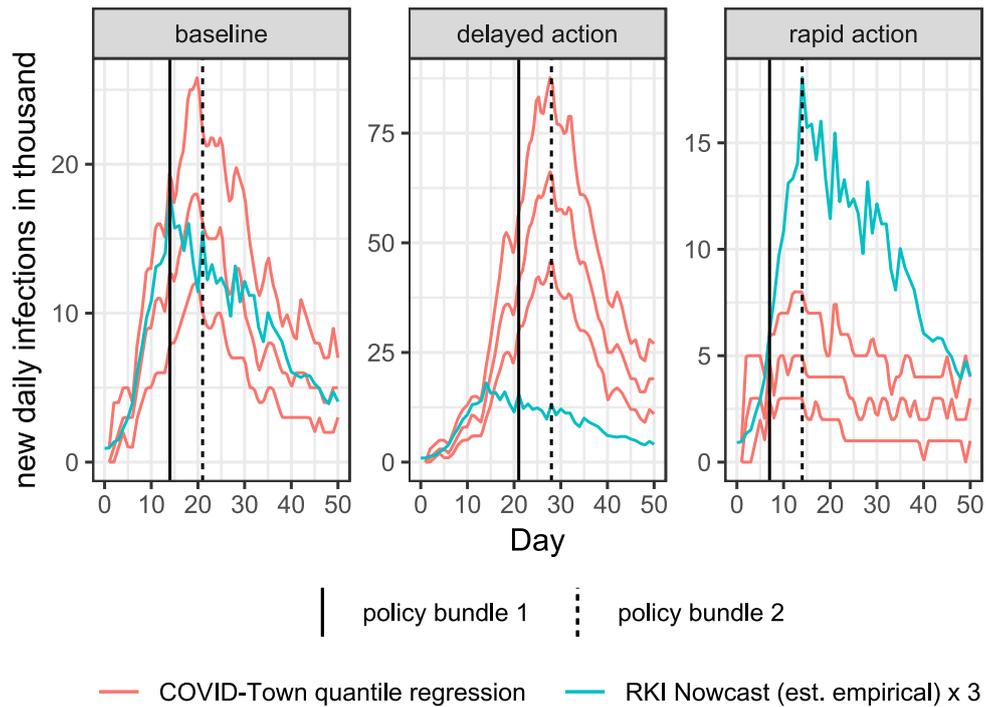

**Figure 7: Infection curves and policy implementation for three epidemiological scenarios**

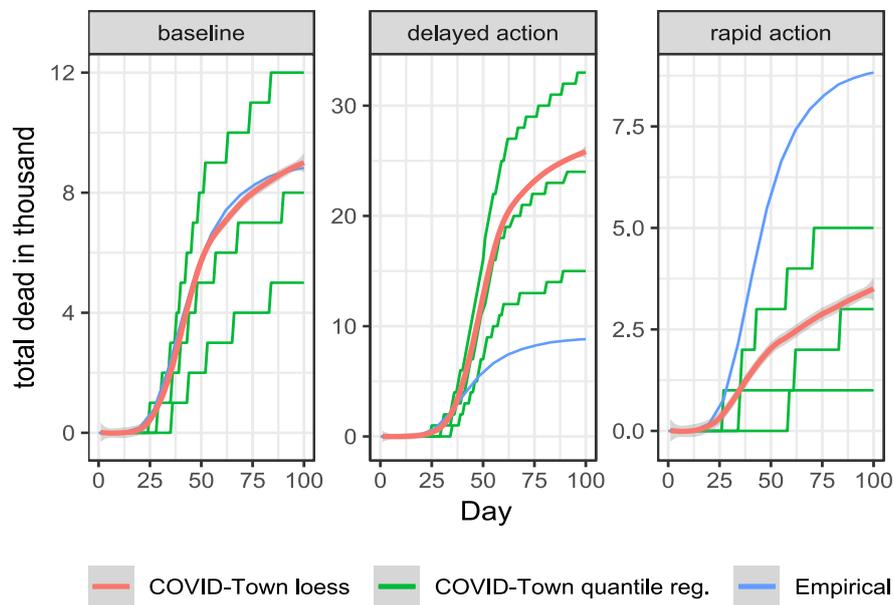

**Figure 8: Deaths for three epidemiological scenarios**

Table 12 shows means and standard deviation for the total dead in thousand and the cumulative consumption goods output lost in percent[16] on day 100 for the three scenarios. The results suggest that rapid action could have saved more than 60% of the lives taken in the first wave, whereas acting one week later would have increased the death toll by more than 180%. Interestingly, the output losses in the consumption goods sector are highest in the delayed action scenario, because it involves more workplace quarantines.

---

[16] This is the difference between the cumulative output that would be expected without the virus (i.e. the production in period 0 multiplied with the number of working days) and actual cumulative output up to period 100.

Table 12: Total dead and output lost for three epidemiological scenarios

|  | baseline | rapid action | delayed action |
|---|---|---|---|
| **Dead in thousand** | 9 (5.656) | 3.493 (2.661) | 25.825 (14.034) |
| **Consumption goods output lost in %** | 3.42 (0.17) | 3.48 (0.09) | 3.49 (0.41) |

Using a Welch test, I find that the differences in mortality, as well as the differences in output between the baseline scenario and the other two scenarios are statistically significant, whereas the output differences between rapid and delayed action are statistically not significant. Looking only at deaths and consumption goods output, delayed action is thus a Pareto-dominated government strategy, even though some groups such as leisure facility owners may benefit from it.

Finally, I test two different stylized fiscal policy scenarios: *Zero-deficit* and *fixed government purchase*, the details of which can be found in subsection 2.5.8. In the zero-deficit scenario, government purchases are limited to current government savings. In the fixed purchase scenario, government purchases are calculated in period zero to match a circular flow of the economy and are then set to be constant throughout the simulation.

In both scenarios, government transfers increase, since the government pays unemployment, sickness, caregiving and quarantine benefits, all of which are bound to increase due to the pandemic and the containment policies. The fixed government purchase scenario thus represents an anticyclical fiscal policy scenario, whereas the zero-deficit scenario is procyclical.

I am interested in both a) the severity of the COVID-19 depression as well as b) the recovery. In order to capture both, I assume that all policies are canceled after 100 days, but since this is only realistic once the virus is eliminated, I only analyze those simulation runs, in which the virus is eliminated within these 100 days, i.e. the opposite condition to the runs shown in figures 7 and 8. These results thus have to be taken with a grain of salt, since the sample only includes the best cases with regard to infections and deaths. Looking only at the recession, however, it does not make a lot of difference to include the other simulation runs. The reason is that only a tiny fraction of the population is infected in any case, which means that output losses due to quarantines and sickness are limited.

Let us first take a look at output in the consumption good sector. In the fixed purchase scenario, output largely drops due to home office orders and school closings, as care giving workers cannot work or work with lower productivity. Once these policies are lifted, output returns to its normal levels very quickly, indicating a V-shaped recovery. In the zero-deficit scenario, on the other hand, the recession quickly worsens after the initial drop, which is the same in both scenarios. There are only small gains after lifting the lockdown policies and even though there is a strong recovery after a while, output still remains flat far below its initial levels thereafter (see fig. 9, left).

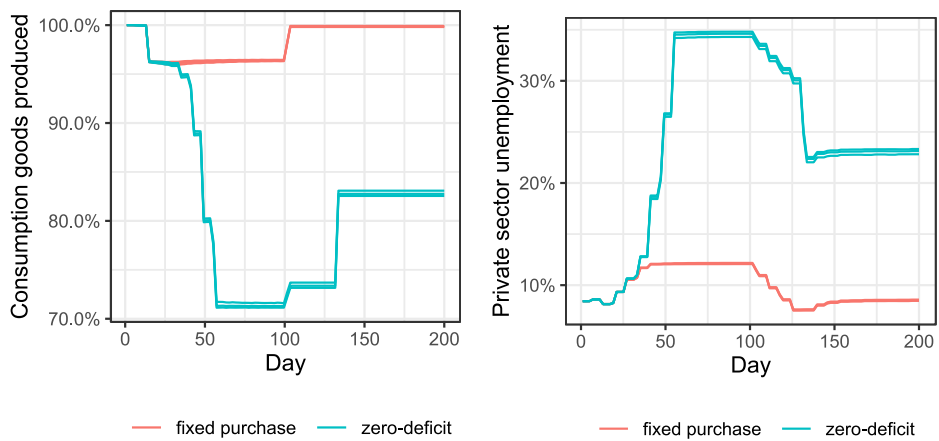

**Figure 9: Output in the consumption good sector (left) and rate of unemployment in the private sector (right)**

Looking at the rate of unemployment in the private sector (see fig. 9, right) delivers a similar picture, even though this figure also depicts the workers in the leisure industry, for whom the recovery takes a longer time. It also uncovers that the recovery in the leisure industry in the zero-deficit scenario ultimately causes the strong (albeit incomplete) recovery later on in the consumption goods industry.

The mechanism that comes into play during the deep recession in the zero-deficit scenario resembles the "Keynesian supply shock" described by Guerrieri et al. (2020). They study the closure of one sector in a two-sector model, where the closure of one sector reduces the income of workers in this sector, which in turn may lower the demand in the other sector, causing production to decline in this sector as well. In my model, however, the transmission mechanism is even graver, because I model the government behavior explicitly. A vicious cycle emerges, in which:

1.) Households cannot spend the money they reserved for leisure activities and thus save it for the time after the relaxation of the measures.

2.) Although the incomes of workers in the leisure industry are secured in the first phase via quarantine benefits, the income of firm owners declines drastically, which in turn translates into reduced consumption.

3.) Tax revenue declines because of the income losses in the leisure industry.

4.) Government expenditure on transfers increases due to the quarantine benefits.

5.) Lower tax revenue and higher government transfers both reduce government purchases.

6.) Lower private and public consumption translates into lower demand in the consumption good sector.

7.) The decreased profitability in the consumption good sector causes offices and factories to fire workers, which lowers tax revenue and increases transfers etc.

Fig. 10 shows the daily leisure activities thwarted, i.e. the sum of leisure plans that failed per day. While the scenarios do not differ until day 100 (left), the differences in income indicated by fig. 9 translates into an increased number of thwarts due to the lack of leisure savings, although the difference between the scenarios becomes smaller over time as leisure facilities adapt their prices. Another reason for the increased number of thwarts can be found in the increased number of unemployed agents who have more leisure periods available per week than employed agents.

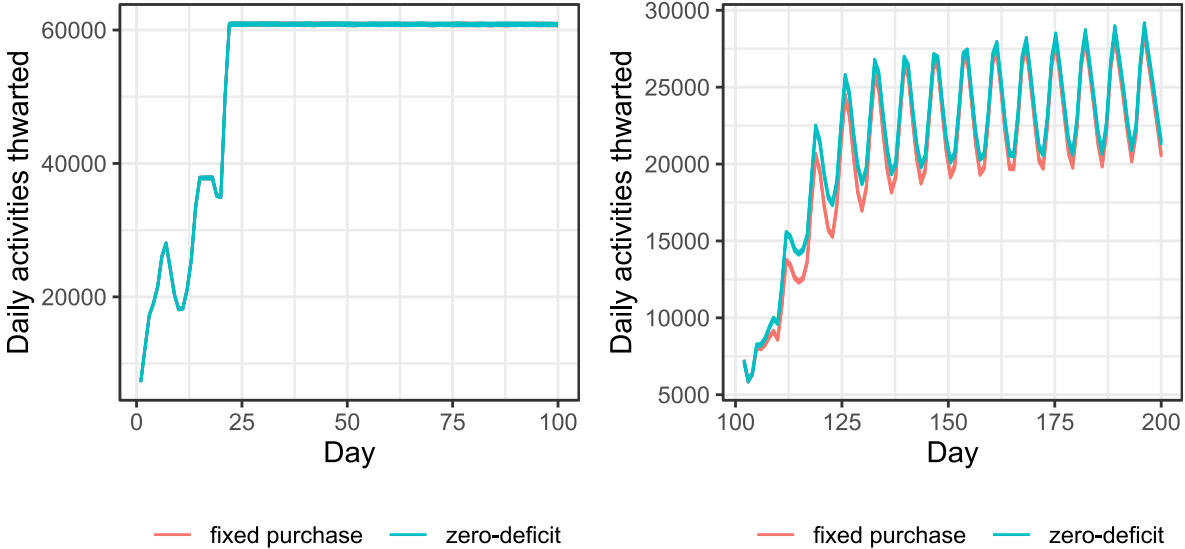

**Figure 10: Daily activities thwarted up to day 100 (left) and after day 100 (right). The figures were split in order to highlight the differences after the end of the containment measures.**

Finally, figure 11 investigates the price at which stabilizing the economy via deficit spending comes. The left side of the graph shows that this policy causes a higher government deficit during the recession, although the deficit is almost recovered after 100 days of recovery. The right side of the graph shows the total number of deaths in each scenario. Interestingly, the increased economic activity in my model does not cause a statistically significant difference in total deaths. Unemployed workers cannot spread the virus at work, but this effect is offset by an increase in contagious leisure activities. Please note that the government also runs a deficit in the "zero-deficit" scenario, since it is assumed to only cut back on government purchases, but not on transfers, if government funds are negative.

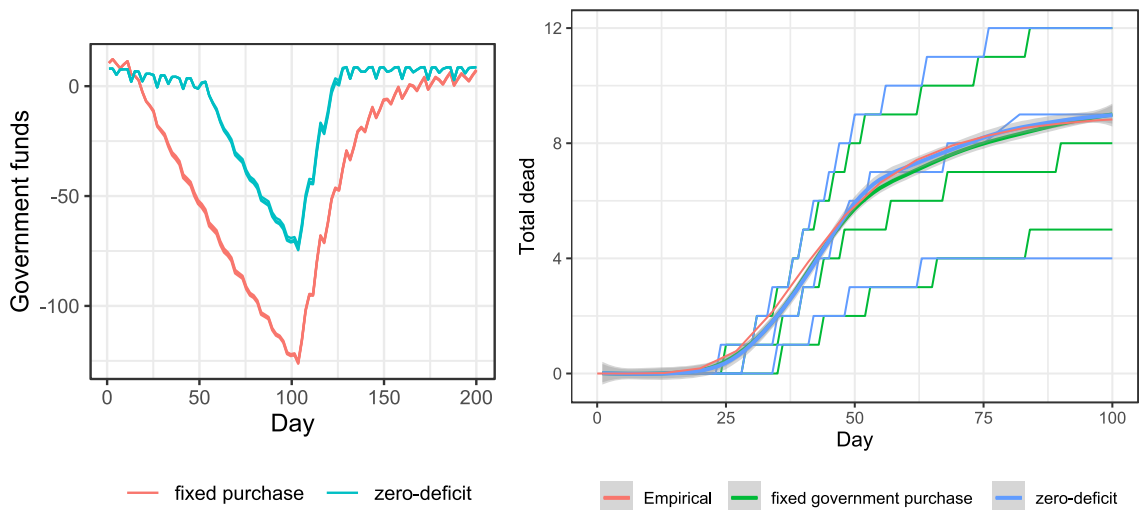

**Figure 11: Government funds in thousand (the daily gross wage of a service worker is the numeraire)**

Table 13 finally shows the number of deaths and the cumulative consumption goods output lost in percent compared to a situation in which no virus and containment measures existed. The first two columns show the situation in period 100 if the virus is not eliminated. The last two columns describe the results for a situation in period 200 where the virus was eliminated by period 100.

**Table 13: Total dead and output lost for two fiscal policy scenarios**

|  | Fixed purchase (not eliminated, after 100 days) | Zero-deficit (not eliminated, after 100 days) | Fixed purchase (eliminated, after 200 days) | Zero-deficit (eliminated, after 200 days) |
|---|---|---|---|---|
| **Dead in thousand** | 9 (5.656) | 8.993 (5.407) | 5.660 (4.398) | 5.962 (4.587) |
| **Consumption goods output lost in %** | 3.42 (0.17) | 18.44 (0.77) | 1.72 (0.10) | 21.68 (1.02) |

A Welch test supports our graphical analysis by a) not finding a statistically significant difference between the two policy scenarios regarding the number of deaths and b) finding a strong difference in cumulative consumption goods output lost. Interestingly, the incomplete recovery is responsible for the fact that cumulative output lost in percent is even worse 100 days after lifting the restrictions than before in the zero-deficit scenario. In the fixed purchase scenario, on the other hand, almost 50% of the lost ground was regained.

## 8 Discussion

Although the current version of COVID-Town captures many stylized facts that conventional SIR-type models are unable to incorporate, many aspects of the model obviously only reflect reality in a stylized way. For instance, the economy in this model is closed, i.e. there is no foreign trade and it only includes three private industries. In order to make e.g. better quantitative predictions about the economic consequences of the pandemic or specific policies, one has to either improve this model's complexity in the relevant areas in order to capture the economic inputs and processes at stake or use this model in conjunction with others.

The spread of the disease could be modeled more realistically by making the locations in this model spatially explicit, as Bicher et al. (2020) and Wallentin et al. (2020) do using GIS data. The distance between the locations then could influence the behavior of the agents (especially leisure preferences). A radical version of this approach was chosen by Bicher et al. (2020), who compute social networks and random interactions primarily based on the spatial location of the agents. Using location data and data on the modal split, one could also model the transmission of the virus via public transportation systems by assuming e.g. that people who use public transportation and commute along a similar route can infect each other.

A more realistic depiction of the economy involves different types of goods, which are sold in different types of shops. For such a model, a lockdown curbs demand for certain types of shops, which in turn puts the related industries into troubles. A shop mechanic involving heterogeneous goods was already implemented by Basurto et al. (2020). However, further room for scientific contributions is left to studying a) cross-elasticities of demand in different industries (e.g. an industry producing home entertainment can be expected to benefit from a prolonged closure of the leisure industry) and b) interconnections between industries from which we can expect nontrivial results that could help to further facilitate the understanding of how the economic recession spreads from one industry to the other.

Another opportunity to improve the model's complexity is given by the pricing mechanism. So far, the consumption goods market is cleared every day. During the beginning of the pandemic, however, everybody witnessed that demand exceeded supply in certain goods deemed essential like toilet paper, pasta or disinfectants and that the pricing mechanism at the same time did not react quickly enough to equalize supply and demand. Introducing an alternative mechanism – as it is common to macroeconomic ABMs (see Dawid and Delli Gatti 2018) – in which firms set their prices and adapt them continuously based on their performance could help to capture this stylized fact and possibly more.

Further questions, such as the role of technology adoption in and technological change after the pandemic, also seem to open up exciting paths of research for agent-based economics, since (macro)economic ABMs can play out their tested strengths (e.g. Lamperti et al. 2018) in a new field of study.

## 9 Conclusion

I developed an agent-based model that allows to study economic and epidemiological consequences of policy scenarios in the COVID-19 pandemic simultaneously and calibrated it to a representative German town. I showed that the model can fit the empirical deaths and estimated infections of the first COVID-19 wave in Germany. In addition to that, the model is able to capture a wide array of stylized empirical facts regarding inputs, processes and outputs of the simulation.

After establishing a validated baseline scenario, I tested two alternative containment scenarios. Monte Carlo simulations suggest that by acting one week earlier, the German government could have saved more than 60% of the lives taken from the beginning of March until the beginning of June 2020. If, on the other hand, the policies are implemented a week later than they actually were, the death toll in COVID-Town grows by more than 180%.

I finally tested two alternative stylized fiscal policy scenarios. Monte Carlo simulations show that the economic fallout of the containment measures can be mitigated with an anticyclical fiscal policy which makes a V-shaped recovery upon lifting the restrictions possible, once the virus is eliminated. If, on the other hand, government purchases are constrained due to financing problems or a balanced budget clause, a vicious cycle emerges, in which private and public spending fall and pull the economy into a deep recession. In this scenario, the recovery after eliminating the virus and lifting the restrictions will be incomplete, marking an L-shaped recovery. Interestingly, the positive economic effects of anticyclical fiscal policy are not coupled with negative epidemiological effects, as there is no statistically significant increase in the number of deaths. This can be explained by the fact that workers who would have worked from home or in workplaces with low or medium infection risks meet infectious individuals during leisure activities.

In addition to the model's focus on economic and epidemiological outcomes, COVID-Town also explicitly models leisure behavior and the need to provide care to young children who cannot go to their school or childcare facility. None of these mechanisms are considered by main economic epidemiological models, but are crucial in the final results of COVID-Town: Care givers cannot work or telework with lower productivity, causing output to drop. Leisure is an important source of infection which offsets counteracts the positive containment effects due to lower economic activity in low or medium contact intensive industries.

The results presented in this paper indicate that an optimal policy mix aiming at containment or holding out for a virus combines rapid containment action to keep the number of infected low with a strong fiscal stimulus to keep employment in industries with low contact intensities up and secure a quick economic recovery after the end of the containment measures.

COVID-Town can be parametrized to other countries and is highly flexible in incorporating and studying the effects of alternative containment scenarios. In the future, the model can also be improved to further investigate the interplay between the economy and the pandemic.

**10 Declarations of interest**: none

**11 Acknowledgements:** I am grateful to my advisor Richard Sturn for supporting my decision to pursue this stream of research. I thank Christian Gehrke and Marcell Göttert for useful comments on previous versions of this paper. I thank Jasper Hepp for making me aware of the model by Basurto et al. (2020) and answering my questions about it. All remaining errors are mine.

**Appendix A**: Sensitivity analysis of the baseline infection probability

Figures A1 and A2 show how the infection and death curves shift, if the baseline infection probability is altered. A higher infection probability would imply a higher dark figure and vice versa.

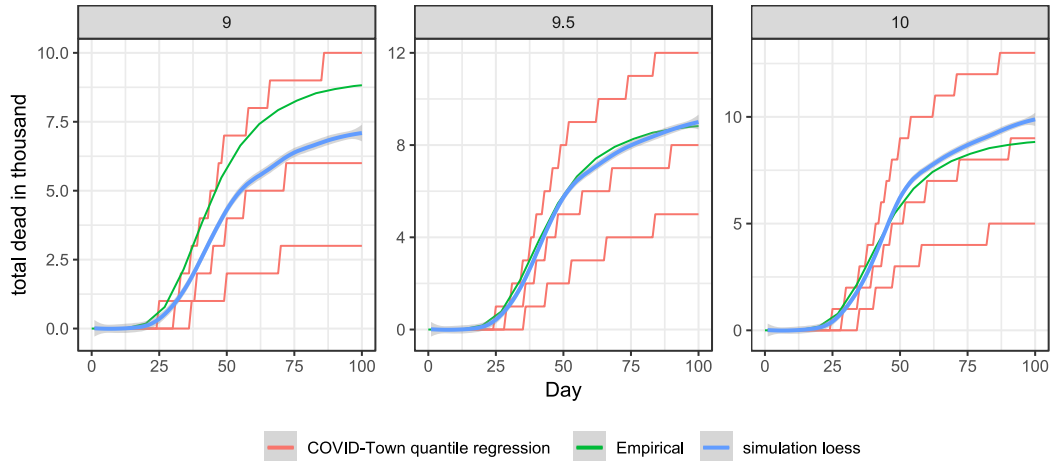

**Figure A 1: Total dead with a baseline infection probability of 9, 9.5 and 10 according to quantile regression and LOESS vs. the empirically observed deaths.**

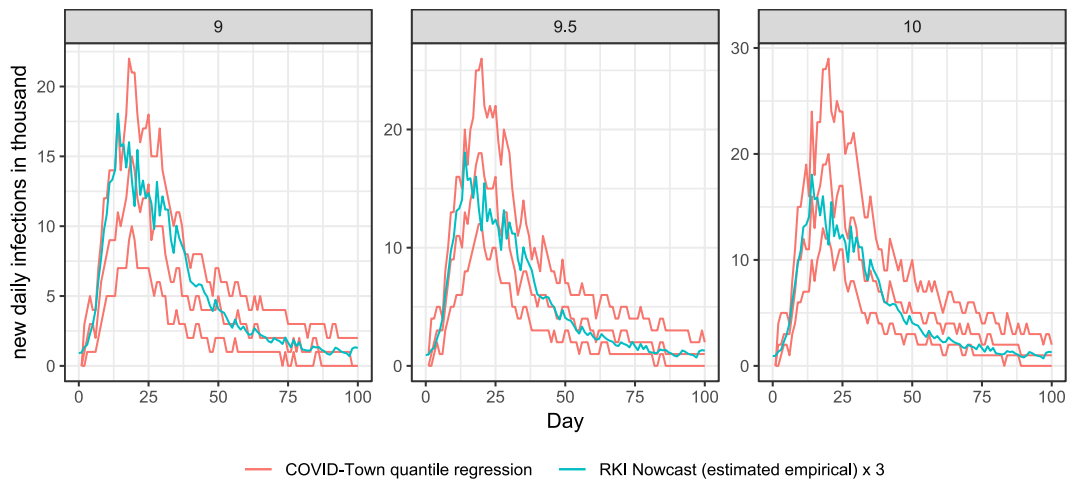

**Figure A 2: New daily infections with baseline infection probability of 9, 9.5 and 10 vs. the dark figure adjusted RKI Nowcast.**

**Appendix B**: Sensitivity analysis for the profit rate buffer

Figure B1 shows that a lower profit rate buffer implies a quicker fall in output, but also a quicker and more complete recovery in the zero-deficit scenario. Although the differences between 0.1 (the baseline parameter) and 0.075 or 0.125 are considerable with regard to the extent of the recovery, the differences between 0.05 and 0.075 or 0.125 and 0.15 are very small or even non-existent.

In the fixed purchase scenario, the profit rate buffer parameter is insensitive to increases. For lower parameter settings, however, the post-lockdown recovery causes a boom and a subsequent recession. This effect can be traced back to the anticyclical nature of the fixed purchase scenario: due to increased employment, the government injects less money into the circular flow via unemployment benefits and extracts more via taxes, thus the economy again. If the profit rate buffer is set to be very low (not shown in this figure), the economy is unable to find an equilibrium as firms constantly hire or fire workers in each week.

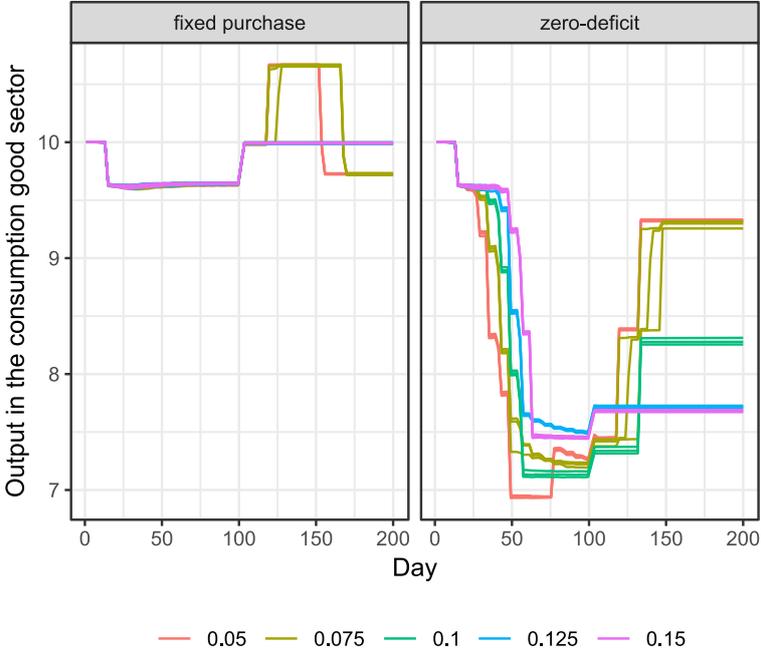

**Figure B 1: Output in the consumption good sector under varying policy scenarios and profit rate buffers.**

**Appendix C**: Sensitivity analysis of the leisure parameters

Figures C1 and C2 show the money that agents saved in total for leisure activities over the first 100 days for a town of 1000 people not exposed to any infection or containment policy. This aggregate measure is important to study the amount of money that is in the circular flow of the economy. While the results do not seem to be very sensitive to changes in the standard deviation of attractiveness of leisure facilities or leisure preferences of the individuals (see fig. C1), fig. C2 shows that the results do depend on the leisure money splash parameter $\omega$ and on the commercial leisure facility attractiveness multiplier $\kappa$. This is intuitive, since both relatively low preferences for visiting commercial leisure facilities and relatively low spending per visit will cause the individuals to spend less money on leisure in total. Both figures also show that the standard parametrization as described in the paper will cause the money saved for leisure to hover around its setup state after 100 days.

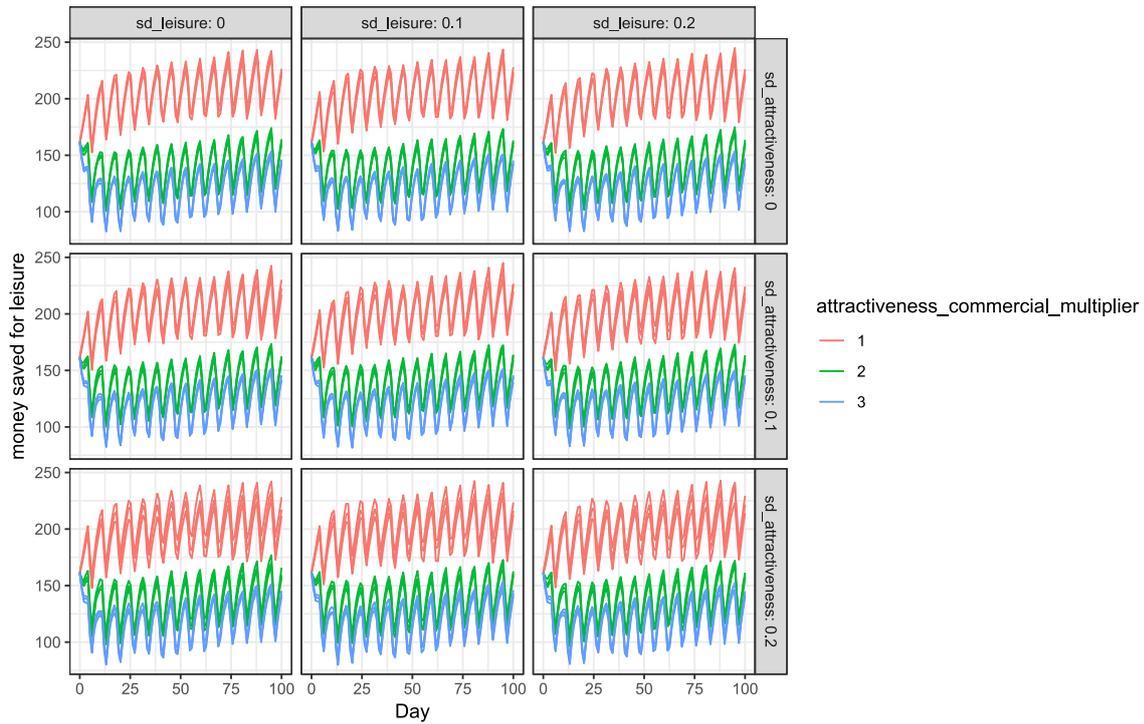

**Figure C 1: Money saved for leisure for different standard deviations for leisure preferences of individuals and attractiveness of facilities, as well as commercial leisure facility attractiveness multipliers.**

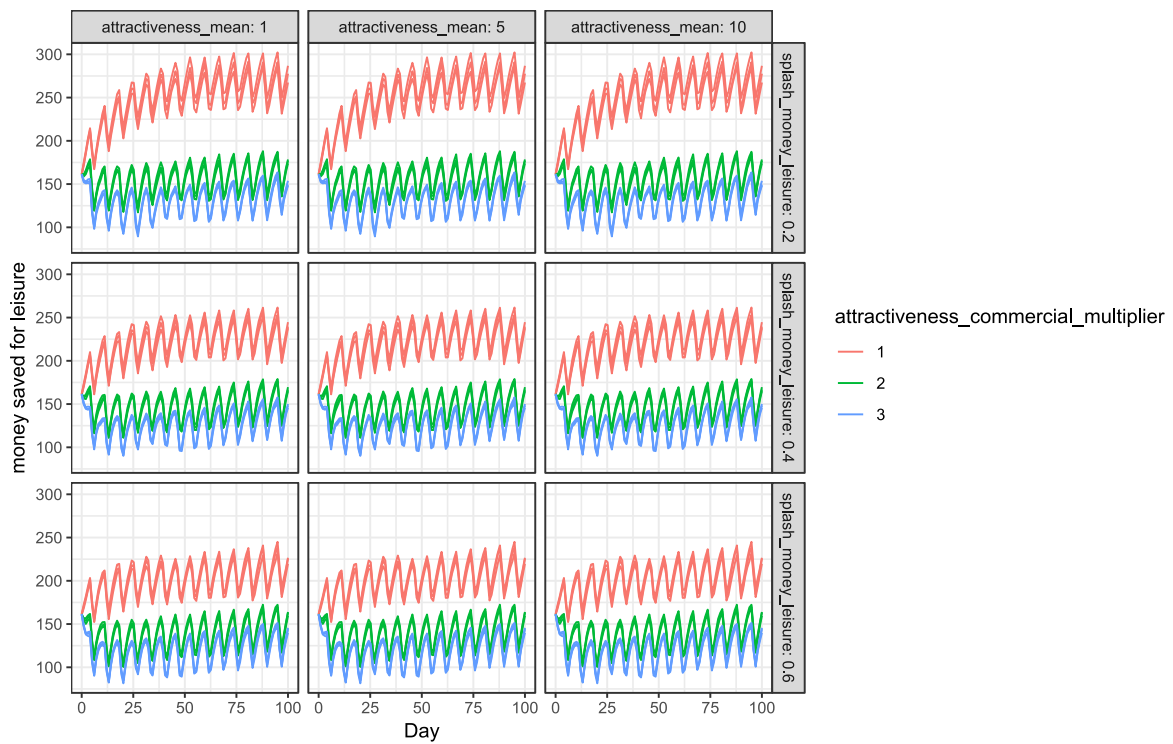

.

**Figure C 2: Money saved for leisure for mean attractiveness of facilities, splash money parameters and commercial leisure facility attractiveness multipliers.**